\documentclass[%
 reprint,
 amsmath,amssymb,
 aps,
]{revtex4-2}

\usepackage{graphicx}
\usepackage{dcolumn}
\usepackage{bm}

\usepackage{graphicx}
\usepackage{epstopdf, epsfig}
\usepackage{lipsum}
\usepackage{amsmath,amssymb}
\usepackage{amsfonts}
\usepackage{graphicx}
\usepackage{todonotes}
\usepackage{tikz}
\usepackage{color}
\usetikzlibrary{shapes}

\definecolor{MATLABcyan}{RGB}{77, 190, 238}
\definecolor{MATLABred}{RGB}{217, 83, 25}
\definecolor{MATLABgreen}{RGB}{119, 172, 48}
\definecolor{MATLAByellow}{RGB}{237, 177, 32}
\definecolor{MATLABpurple}{RGB}{126, 47, 142}
\definecolor{MATLABblue}{RGB}{0, 114, 189}
\definecolor{MATLABgrey}{RGB}{128, 128, 128}

 \newcommand{\Mgreen}{\raisebox{0.5pt}{\tikz{\node[draw=MATLABgreen,scale=0.5,circle,fill=MATLABgreen](){};}}}
 \newcommand{\Mred}{\raisebox{0.5pt}{\tikz{\node[draw=MATLABred,scale=0.5,circle,fill=MATLABred](){};}}}
\newcommand{\Myellow}{\raisebox{0.5pt}{\tikz{\node[draw=MATLAByellow,scale=0.5,circle,fill=MATLAByellow](){};}}}
\newcommand{\Mcyan}{\raisebox{0.5pt}
{\tikz{\node[draw=MATLABcyan,scale=0.5,circle,fill=MATLABcyan](){};}}}
\newcommand{\Mblue}{\raisebox{0.5pt}
{\tikz{\node[draw=MATLABblue,scale=0.5,circle,fill=MATLABblue](){};}}}
\newcommand{\Mpurple}{\raisebox{0.5pt}{\tikz{\node[draw=MATLABpurple,scale=0.5,circle,fill=MATLABpurple](){};}}}

\begin{document}

\preprint{APS/123-QED}

\title[]{Active particle motion in Poiseuille flow through rectangular channels}

\author{Rahil N. Valani$^{1,3}$}\email{rahil.valani@physics.ox.ac.uk}
\author{Brendan Harding$^{2}$}
\author{Yvonne M. Stokes$^{1}$}%
\affiliation{$^1$School of Computer and Mathematical Sciences, University of Adelaide, South Australia 5005, Australia}
\affiliation{$^2$School of Mathematics and Statistics, Victoria University of Wellington, Wellington 6012, New Zealand}
\affiliation{$^3$Rudolf Peierls Centre for Theoretical Physics, Parks Road,
University of Oxford, OX1 3PU, United Kingdom}

\date{\today}

\begin{abstract}

We investigate the dynamics of a point-like active particle suspended in fluid flow through a straight channel. For this particle-fluid system, we derive a constant of motion for a general unidirectional fluid flow, and apply it to an approximation of Poiseuille flow through channels with rectangular cross-sections. We obtain a $4$D nonlinear conservative dynamical system with one constant of motion and a dimensionless parameter describing the ratio of maximum flow speed to intrinsic active particle speed. Applied to square channels, we observe a diverse set of active particle trajectories with variations in system parameters and initial conditions which we classify into different types of swinging, trapping, tumbling and wandering motion. Regular (periodic/quasiperiodic) motion as well as chaotic active particle motion are observed for these trajectories and quantified using largest Lyapunov exponents. We explore the transition to chaotic motion using Poincar\'e maps and show ``sticky" chaotic tumbling trajectories that have long transients near a periodic state. We briefly illustrate how these results extend to rectangular cross-sections with width/height ratio larger than one. Outcomes of this work may have implications for dynamics of natural and artificial microswimmers in experimental microfluidic channels that typically have rectangular cross-sections.

\end{abstract}

\maketitle


\section{\label{sec: intro} Introduction}

Active particles are entities that take energy from the environment and convert it into persistent motion. Examples include macroscopic living organisms, such as birds, fish and mammals, which consume energy from food and self-propel via various modes of locomotion. Active particles are also ubiquitous in the microscopic living world such as bacteria, cells, algae and other microorganisms~\citep{animate1}. Although persistent motion is a visible feature that is commonly associated with life, active particles also emerge in several non-equilibrium inanimate physical and chemical systems~\citep{iunanimate7,inanimate10,Couder2005WalkingDroplets,superwalker,stopgoswim}. 

Active particles immersed in a fluid medium at the micro scale, also known as microswimmers, are a commonly studied class of active particles~\citep{microswim1}. These microswimmers ubiquitously interact with external fluid flows in various situations. For example, microswimmers routinely experience unidirectional flows in confined channels such as sperm cells swimming in fallopian tubes~\citep{doi:10.1073/pnas.1120955109,Simons2018}, pathogens moving through blood vessels~\citep{doi:10.1146/annurev-chembioeng-060817-084006} and micro-robots programmed for targeted drug delivery applications~\citep{D0CS00309C}. In these scenarios, the coupling between external flow fields and intrinsic velocity of the active particle can lead to rich dynamical behaviors~\citep{Microswimmersreview}. Understanding of the active particle dynamics arising from coupling with external fluid flows is not only interesting 
from a biological perspective, but 
is also crucial for design of artificial microswimmers for biomedical applications of cell manipulation, targeted drug delivery and cargo transport~\citep{mi11121048}. Further, it can aid 
design of industrial and biomedical microfluidic devices aimed at focusing, sorting and filtering of microorganisms in a fluid suspension~\citep{Zhang2018,PhysRevLett.119.198002}.

\citet{Zottl2012} studied the motion of a microswimmer in unidirectional confined flows by modeling the active particle as a spherically symmetric point with constant intrinsic velocity. For $2$D planar Poiseuille flow, they showed that the equation of motion for the active particle can be mapped onto the mathematical equation of a simple pendulum, where the oscillating and circling solutions of the pendulum motion correspond to two different types of active particle motion, swinging and tumbling, respectively. In swinging motion the upstream-oriented active particle performs oscillations about the channel centerline, whereas in tumbling motion, the active particle oscillates near the edges of the channel with fluctuating orientation and does not cross the channel centerline. Variations of this model that include additional attributes to the active particle and/or fluid flow have been investigated in detail for $2$D channel flows~\citep{twoDmicroswimmer1,twoDmicroswimmer2,twoDmicroswimmer3,Choudhary2022,twoDmicroswimmer4,ellipactiveparticle}. For $3$D cylindrical Poiseuille flow, \citet{Zottl2012} showed that the particle-fluid dynamical system is Hamiltonian with enough conserved quantities to make the system integrable. In this case, they showed that the active particle exhibits periodic motion with $3$D generalizations of swinging and tumbling trajectories. The effect of flow anisotropy was also studied by \citet{ellipactiveparticle} 
who showed, for 
an elliptical channel cross-section, 
that the active particle motion is much more complex with, typically, quasiperiodic trajectories. 
Using Poincar\'e maps, a few examples of chaotic motion  were also reported by \citet{Zottlthesis}. 

Although the axisymmetric fluid flow profile in a cylindrical channel results in simplified equations for the active particle motion, in many microfluidic applications concerned with natural and artificial microswimmers, microchannels with rectangular cross-sections are more commonly used since they are relatively easy to fabricate~\citep{doi:10.1146/annurev.matsci.28.1.153}. 
Motivated by this, herein we apply the model of \citet{Zottl2012} to explore the dynamics of a simple active particle suspended in Poiseuille flow through a straight channel having square/rectangular cross-section. The introduction of a square/rectangular cross-section introduces anisotropy, by breaking the continuous rotational symmetry of fluid flow that exists in a circular cross-section, and makes the system non-integrable. We observe a rich variety of active particle motion with both quasiperiodic and chaotic trajectories.
These motions are explored in detail as a function of system parameters and initial conditions.

The paper is organized as follows. In Sec.~\ref{sec: eq} we present the equations of motion for the particle-fluid system and derive general constants of motion for an active particle in unidirectional fluid flow. We then, in Sec.~\ref{sec: eq lin stab}, identify equilibrium states for an active particle suspended in Poiseuille-like flow through a rectangular cross-section and determine their stability. 
After briefly reviewing the special case of active particle motion in a channel with circular cross-section in Sec.~\ref{sec: circle}, we present a detailed exploration of active particle dynamics in a channel with square cross-section in Sec.~\ref{sec: square}. This includes a classification of trajectories, comparison with dynamics in a circular cross-section, a detailed parameter space exploration of cross-sectional active particle dynamics, as well as an investigation of the transport of an active particle along the channel. In Sec.~\ref{sec: rect} we briefly explore the effect of the width/height ratio of the rectangular cross-section on active particle motion. We provide our conclusions in Sec.~\ref{sec: concl}.

\begin{figure}
\centering
\includegraphics[width=\columnwidth]{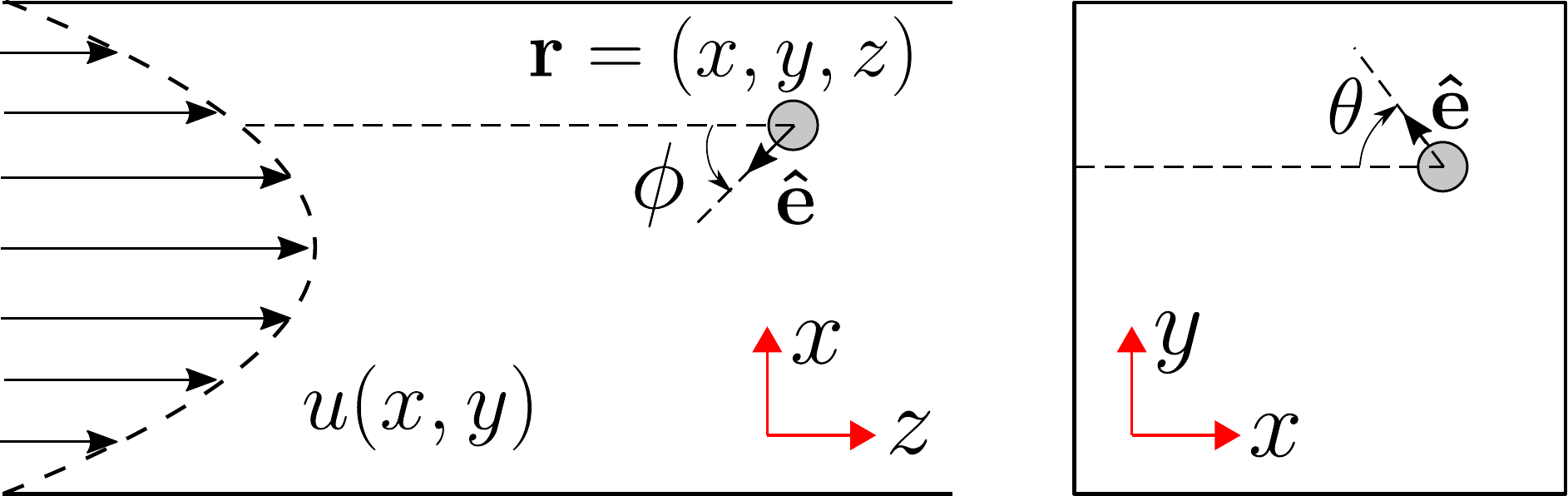}
\caption{Schematic of the particle-fluid system. 
A simple point-like active particle located at $\mathbf{r}=(x,y,z)$ and having a constant 
intrinsic speed in the direction of its orientation $\mathbf{\hat{e}}$ is suspended in a unidirectional channel flow ${u}(x,y)$ through a straight $3$D rectangular channel with width to height aspect ratio $AR$. The particle's orientation is represented using spherical co-ordinates with polar angle $\theta\in(-\pi/2,\pi/2)$ measuring the orientation relative to the $x-z$ plane, and azimuthal angle $\phi\in(-\pi,\pi]$ measuring angle within the $x-z$ plane relative to the (negative) $z$ axis. The left panel shows the top-view of the channel in the $x-z$ plane while the right panel shows the cross-sectional view of the channel in the $y-x$ plane.}
\label{Fig: schematic}
\end{figure}

\section{\label{sec: eq} Equations of motion}

Consider the point-like active particle model of a spherical microswimmer illustrated in Fig.~\ref{Fig: schematic}. The active particle has 
a constant intrinsic swimming speed $v_0$ in the direction of its orientation $\mathbf{\hat{e}}=e_x\mathbf{\hat{i}}+e_y\mathbf{\hat{j}}+e_z\mathbf{\hat{k}}$, 
is located at $\mathbf{r}=x\mathbf{\hat{i}}+y\mathbf{\hat{j}}+z\mathbf{\hat{k}}$, and is suspended in a steady unidirectional flow $u(x,y)\mathbf{\hat{k}}$ through a straight $3$D channel. The equations of motion for the active particle are given by~\citep{Zottl2012}:
\begin{subequations}
\begin{align}\label{eq: dimensional position}
    \frac{\text{d}\mathbf{r}}{\text{d}t}&=v_0\mathbf{\hat{e}}+u(x,y)\mathbf{\hat{k}},
\\\label{eq: dimensional orientation}
    \frac{\text{d}{\mathbf{\hat{e}}}}{\text{d}t}&=\frac{1}{2}(\nabla \times u(x,y)\mathbf{\hat{k}})\times \mathbf{\hat{e}}.
\end{align} 
\end{subequations}
Equation~\eqref{eq: dimensional position} describes the translational motion of the active particle as a combination of its intrinsic velocity $v_0\mathbf{\hat{e}}$ and the local velocity of the background fluid flow $u(x,y)\mathbf{\hat{k}}$, whereas Eq.~\eqref{eq: dimensional orientation} describes the evolution of the active particle's orientation based on the local flow vorticity. We assume that the active particle is small compared to the cross-sectional dimensions of the channel, and hence the particle does not disturb the fluid flow. Further, we assume
that the active particle stays away from bounding walls so
that we can neglect interactions/collisions
between the active particle and the walls. 

Non-dimensionalizing Eqs.~\eqref{eq: dimensional position} and \eqref{eq: dimensional orientation} with a characteristic length scale $H$ of the cross-section and time scale $H/v_0$, we obtain the following dimensionless equations:
\begin{subequations}\label{nonlinear eq AR cartesian full}            
\begin{align}
    \frac{\text{d}\bar{\mathbf{r}}}{\text{d}\bar{t}}&=\mathbf{\hat{e}}+{\bar{u}(\bar{x},\bar{y})\mathbf{\hat{k}}},
\\
    \frac{\text{d}{\mathbf{\hat{e}}}}{\text{d}\bar{t}}&=\frac{1}{2}(\bar{\nabla} \times{\bar{u}}(\bar{x},\bar{y})\mathbf{\hat{k}})\times \mathbf{\hat{e}}.
\end{align}     
\end{subequations}      
Here, the dimensionless variables are denoted with an overbar and the dimensionless flow field ${\bar{u}}(\bar x,\bar y)$ is scaled with the active particle speed $v_0$. We now drop the overbars on dimensionless variables for convenience. In component form, we get a system of six nonlinear ordinary differential equations (ODEs) as follows:
\begin{align*}
   \dot{x}&=e_x, & \dot{e}_x&=-\frac{1}{2} e_z  \frac{\partial {u}}{\partial x},\\
   \dot{y}&=e_y, & \dot{e}_y&=-\frac{1}{2} e_z  \frac{\partial {u}}{\partial y},\\
   \dot{z}&=e_z + {u}(x,y), & \dot{e}_z&=\frac{1}{2} e_x \frac{\partial {u}}{\partial x} + \frac{1}{2} e_y  \frac{\partial {u}}{\partial y}.
\end{align*}
We note from the component form that the dynamical flow is divergence free, that is 
$$\frac{\partial \dot{x}}{\partial x}+\frac{\partial \dot{y}}{\partial y}+\frac{\partial \dot{z}}{\partial z}+\frac{\partial \dot{e}_x}{\partial e_x}+\frac{\partial \dot{e}_y}{\partial e_y}+\frac{\partial \dot{e}_z}{\partial e_z}=0.$$
Hence, the dynamical system is conservative and phase-space volumes are preserved under the dynamical flow. We further note that the $z$ variable can be decoupled, i.e. the $\dot{z}$ equation can be integrated separately, thus reducing our dynamical system to five differential equations. The effective dimension of our dynamical system is further reduced by identifying constants of motion, i.e. quantities that remain constant during the evolution of the system. A trivial constant of motion for our system is 
\begin{equation}\label{eq: COM trivial}
|\mathbf{\hat{e}}|^2=e_x^2+e_y^2+e_z^2=1,
\end{equation}
since the orientation vector maintains unit magnitude. We have also identified a second constant of motion as (see Appendix~\ref{sec: general COM} for a proof) 
\begin{equation}\label{eq: COM general}
H_g=-\frac{1}{2}{u}(x,y)+e_z.
\end{equation}
With these two constants of motion, our five-dimensional dynamical system reduces to three effective dimensions. 
We can implicitly use the constant of motion in Eq.~\eqref{eq: COM trivial} and reduce our system to four nonlinear ODEs by parameterizing the Euler axis using spherical co-ordinate angles $\theta$ and $\phi$ as follows:
\begin{align*}
    e_x&=-\cos\theta\,\sin\phi,\\
    e_y&=\sin\theta,\\
    e_z&=-\cos\theta\,\cos\phi.
\end{align*}
Here, $\theta\in(-\pi/2,\pi/2)$ is the polar angle measuring the orientation relative to the $x-z$ plane, while $\phi \in (-\pi,\pi]$ is the azimuthal angle measuring the orientation component within the $x-z$ plane relative to the negative $z$ axis~(see Fig.~\ref{Fig: schematic}).
This parameterization gives us the following four coupled nonlinear ODEs:
\begin{subequations}\label{nonlinear eq AR gen}
\begin{align}
    \dot{x}&=-\cos\theta \sin\phi,\\ 
    \dot{y}&=\sin\theta,\\ 
    \dot{\theta}&=\frac{1}{2}\frac{\partial {u}}{\partial y}\cos\phi,\\ 
    \dot{\phi}&=\frac{1}{2}\frac{\partial {u}}{\partial y}\tan\theta\,\sin\phi-\frac{1}{2}\frac{\partial {u}}{\partial x},
\end{align}
\end{subequations}
along with the 
constant of motion in Eq.~\eqref{eq: COM general} which is rewritten in the above parameterization as
\begin{equation}\label{eq: COM general spehrical}
H_g=-\frac{1}{2}{u}(x,y)-\cos\phi \cos\theta.
\end{equation}
We note that up to this point, our consideration of the fluid flow field ${u}(x,y)$ has been general and hence the constant of motion in Eq.~\eqref{eq: COM general} exists independent of the specific flow profile.

We now consider the specific fluid flow profile of Poiseuille flow in a $3$D straight channel having a rectangular cross-section with width $W$ and height $H$. With $H$ as the length scale and defining the aspect ratio $AR=W/H$, we approximate the dimensionless flow profile by 
\begin{equation}\label{eq: flow field dimless}
{u}(x,y)=U\left(1-\left(\frac{x}{AR}\right)^2\right)\left(1-y^2\right),
\end{equation}
where $U=u(0,0)$ is the maximum velocity in the channel scaled with the intrinsic particle speed $v_0$.
This expression provides a good approximation to the exact solution of Poiseuille flow in rectangular channels expressed as an infinite series~(see Appendix~\ref{sec: flow compare}). 
Substituting this flow field in Eq.~\eqref{nonlinear eq AR gen} we get the following $4$D dynamical system (along with the constant of motion in Eq.~\eqref{eq: COM general spehrical}):
\begin{subequations}\label{nonlinear eq AR}
\begin{align}
    \dot{x}&=-\cos\theta \sin\phi,\\ 
    \dot{y}&=\sin\theta,\\ 
    \dot{\theta}&=-{U}y\cos\phi\left(1-\frac{x^2}{AR^2}\right),\\ 
    \dot{\phi}&=-{U}y\tan\theta \sin\phi \left(1-\frac{x^2}{AR^2}\right) + {U}\frac{x(1-y^2)}{AR^2},
\end{align}
\end{subequations}
where $-AR< x <AR$, $-1<y<1$, $-\pi/2< \theta < \pi/2$ and $-\pi<\phi\leq\pi$. 

We solve the dynamical system in Eq.~\eqref{nonlinear eq AR} up to $t=1000$ (unless stated otherwise) using the ode45 solver in MATLAB with relative and absolute tolerance of $10^{-10}$. These very small tolerances ensure that numerical variations in the constants of motion are less than $10^{-8}$ for the duration of simulations.

\section{Equilibrium states and stability}\label{sec: eq lin stab}

We start by finding equilibrium states of the dynamical system that would correspond to an active particle with a fixed cross-sectional location and a fixed orientation. This is done by making the time derivatives zero in Eq.~\eqref{nonlinear eq AR} and solving the resulting nonlinear algebraic equations. We find the following two equilibrium states:
$$(x^*,y^*,\theta^*,\phi^*)=(0,0,0,0)\,\,\text{and}\,\,(0,0,0,\pi).$$
The first equilibrium (with $\phi^*=0$) 
corresponds to an active particle oriented upstream at the center of the channel, while the second (with $\phi^*=\pi$) 
corresponds to an active particle oriented downstream, also at the center of the channel.

To understand the stability of these equilibrium states, we perform a linear stability analysis, perturbing the equilibrium states thus: $(x,y,\theta,\phi)=(x^*,y^*,\theta^*,\phi^*)+\epsilon (x_1,y_1,\theta_1,\phi_1)$, where $0<\epsilon\ll 1$ is a perturbation parameter. Substituting this in Eq.~\eqref{nonlinear eq AR} and comparing $O(\epsilon)$ terms we get a matrix equation for the evolution of the perturbation variables $(x_1,y_1,\theta_1,\phi_1)$.  

\subsubsection{Upstream-oriented equilibrium state}\label{sec: upstream eigenvalues}

For the upstream-oriented equilibrium state $(x^*,y^*,\theta^*,\phi^*)=(0,0,0,0)$, we obtain the following linear equation that governs the evolution of perturbations:
\begin{gather*}
 \begin{bmatrix} 
 \dot{x}_1 \\
 \dot{y}_1 \\
 \dot{\theta}_1 \\
 \dot{\phi}_1 
 \end{bmatrix}
 =
  \begin{bmatrix}
0 & 0 & 0 & -1\\
0 & 0 & 1 & 0\\
0 & -{U} & 0 & 0\\
{U}/AR^2 & 0 & 0 & 0
 \end{bmatrix}
  \begin{bmatrix}
  {x}_1 \\
 {y}_1 \\
 {\theta}_1 \\
  {\phi}_1 \\
 \end{bmatrix}.
\end{gather*}
The stability of the equilibrium state is determined by the nature of the eigenvalues of the right-hand-side matrix~\citep{strogatz}. We obtain the following characteristic polynomial equation for eigenvalues $\lambda$:
$$\lambda^4+{U}\left(1+\frac{1}{AR^2}\right)\lambda^2 +\frac{{U}^2}{AR^2}=0.$$
The roots of this quartic polynomial give the eigenvalues
$$\lambda=\pm i \sqrt{{U}},\: \pm i \frac{\sqrt{{U}}}{AR}.$$
All the eigenvalues being purely imaginary, this upstream-oriented active particle equilibrium may correspond to a center or a stable/unstable spiral~\citep{strogatz}. However, the conservative nature of our dynamical system makes this equilibrium a center. We will revisit this numerically in Sec.~\ref{sec: motion eq}. Furthermore, for non-square cross-sections having 
aspect ratio 
differing from $AR=1$, the oscillation frequency differs along the two eigenvector pairs corresponding to the two conjugate eigenvalue pairs.

\begin{figure*}
\centering
\includegraphics[width=2\columnwidth]{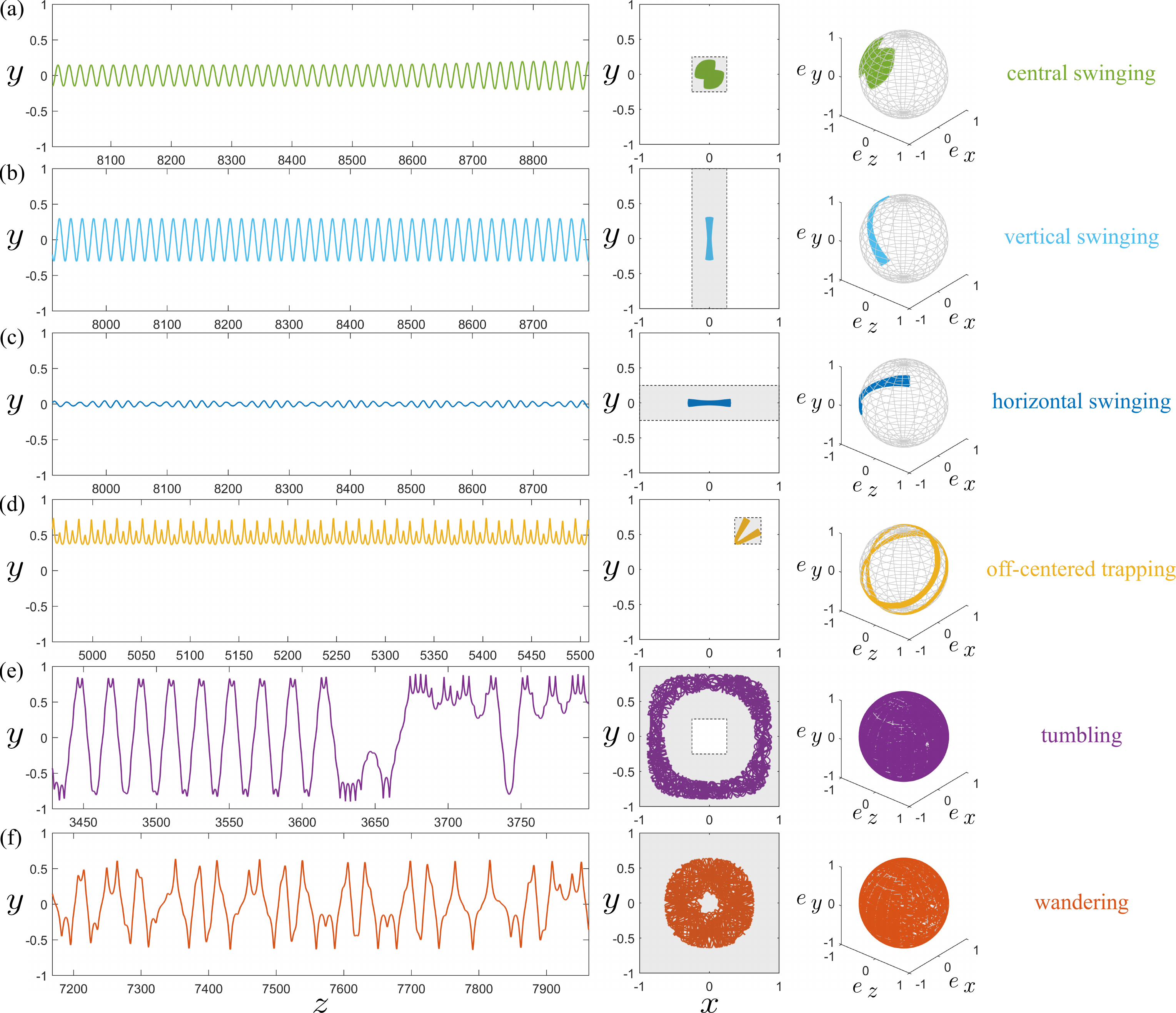}
\caption{Classification of active particle motion. Different trajectories of the active particle in a straight channel with square cross-section are shown in the $(z,y)$ plane and $(x,y)$ plane based on different initial positions for: (a) $(x(0),y(0))=(0.15,0.15)$, (b) $(x(0),y(0))=(0.05,0.30)$, (c)  $(x(0),y(0))=(0.30,0.05)$, (d)  $(x(0),y(0))=(0.58,0.70)$, (e) $(x(0),y(0))=(0.75,0.75)$ and (f)  $(x(0),y(0))=(0.55,0.40)$. At right, the shaded region of the spherical surface shows the time evolution of the particle's orientation.
The different colors represent the classification of trajectory based on the region occupied in the cross-section. (a) Central swinging motion (green); trajectories that stay near the center of the channel inside the gray square $-0.25<x,y<0.25$. (b) Vertical swinging motion (cyan); trajectories that are confined in the $x$ direction i.e. inside the vertical band $-0.25<x<0.25$. (c) Horizontal swinging motion (blue); trajectories that are confined in the $y$ direction i.e. inside the horizontal band $-0.25<y<0.25$. (d) Off-centered trapping (yellow); trajectories that are confined in a rectangular region of the cross section having an area less than half the area of the cross-section and which may cross either centreline ($x=0$, $y=0$), or neither, but not both. 
(e) Tumbling motion (purple); trajectories that stay near the walls of the channel, i.e. outside the region of central swinging motion.  (f) Wandering motion (red); trajectories that explore both the central as well as outer regions of the cross-section. Other parameters were fixed to ${U}=10$, $z(0)=0$, $\theta(0)=0$ and $\phi(0)=0$. See also supplemental videos S1-S6 for videos of particle trajectories and orientations corresponding to panels (a)-(f), respectively.}
\label{Fig: types of traj}
\end{figure*}

\subsubsection{Downstream-oriented equilibrium state}

For the downstream-oriented equilibrium state $(x^*,y^*,\theta^*,\phi^*)=(0,0,0,\pi)$, we obtain the following linear equation describing the evolution of perturbations:
\begin{gather*}
 \begin{bmatrix} 
 \dot{x}_1 \\
 \dot{y}_1 \\
 \dot{\theta}_1 \\
 \dot{\phi}_1 
 \end{bmatrix}
 =
  \begin{bmatrix}
0 & 0 & 0 & 1\\
0 & 0 & 1 & 0\\
0 & {U} & 0 & 0\\
{U}/AR^2 & 0 & 0 & 0
 \end{bmatrix}
  \begin{bmatrix}
  {x}_1 \\
 {y}_1 \\
 {\theta}_1 \\
  {\phi}_1 \\
 \end{bmatrix},
\end{gather*}
with the characteristic equation
$$\lambda^4-{U}\left(1+\frac{1}{AR^2}\right)\lambda^2 +\frac{{U}^2}{AR^2}=0.$$
Solving this quartic polynomial gives us the eigenvalues
$$\lambda=\pm \sqrt{{U}},\: \pm \frac{\sqrt{{U}}}{AR}.$$
Since the eigenvalues are all real, with two positive and two negative, this equilibrium state is an unstable saddle point having both stable and unstable manifolds which are two dimensional. 
Again, due to the different magnitudes of eigenvalues for non-square cross-sections, the rate of instability from the saddle point differs in the directions of the corresponding eigenvectors. We will numerically revisit the dynamics of an active particle starting near this equilibrium state in Sec.~\ref{sec: motion eq}.

\section{\label{sec: circle} Dynamics in a cylindrical channel}

\citet{Zottl2012} studied the nonlinear dynamics of an active particle suspended in fluid flow through a cylindrical channel.
For a circular cross-section, the flow field is axisymmetric and takes the dimensionless form
\begin{equation}\label{eq: flow circular}
u_{c}(x,y)={U}\left(1-(x^2+y^2)\right),
\end{equation}
(with length variables scaled with the duct radius).
For this special case, they identified two constants of motion:
\begin{subequations}
\begin{align}\label{eq: COM circle 1}
H_c&=\frac{1}{2}{U}(x^2+y^2)+1-\cos\phi \cos\theta,    
\\
\label{eq: COM circle 2}
L_z&=x \sin\theta + y \cos\theta \sin\phi.
\end{align}
\end{subequations}
Here $H_c$ is a linear transform of the general constant of motion identified in Eq.~\eqref{eq: COM general spehrical}. The new constant of motion $L_z$ arises from the continuous rotational symmetry of the circular cross-section and it is proportional to the angular momentum of the active particle in the $z$ direction. This additional constant of motion further reduces the effective dimension of the dynamical system in Eq.~\eqref{nonlinear eq AR gen} from three to two dimensions. By making a change of co-ordinates, \citet{Zottl2012} obtained three nonlinear ODEs with the two constants of motion $H_c$ and $L_z$, and showed that the active particle motion in a cylindrical channel results in an integrable Hamiltonian system where the motion in the three-dimensional phase-space is restricted to a curve formed by the intersection of two surfaces corresponding to $H_c$ and $L_z$.

\section{\label{sec: square} Dynamics in a square channel}

We now explore in detail the dynamics of an active particle suspended in fluid flow through a $3$D straight channel with a square cross-section i.e. $AR=1$. We explore the active particle dynamics as a function of the dimensionless parameter ${U}$ as well as the initial conditions i.e. the initial position in the cross-section $(x(0),y(0))$, and the initial orientation angles $(\theta(0),\phi(0))$.

\subsection{\label{sec: traj} Classification of active particle trajectories}

A large diversity of trajectories are observed for the active particle in a square channel by varying the system parameter ${U}$ as well as the initial conditions. Some typical trajectories and the corresponding orientations are shown in Fig.~\ref{Fig: types of traj} for ${U}=10$. We choose to classify the trajectories into the following six types based on the region they occupy (shown in gray in Fig.~\ref{Fig: types of traj}) at long times within the square cross-section. 
(i) Central swinging motion (green) with trajectories undergoing swinging motion about the channel centerline (similar to swinging motion in cylindrical channel~\citep{Zottl2012}) and confined near the center of the channel 
to the cross-sectional domain $-0.25<x,y<0.25$. (ii) Vertical swinging motion (cyan) with trajectories undergoing swinging motion in the vertical $y$ direction and confined in the $x$ direction 
to the vertical band $-0.25<x<0.25$. (iii) Horizontal swinging motion (blue) with trajectories undergoing swinging motion in the horizontal $x$ direction and confined in the $y$ direction to 
the horizontal band $-0.25<y<0.25$. (iv) Off-centered trapping (yellow) with trajectories confined within a rectangular region of the cross-section having an area less than half the area of the cross-section; this region may cross at most one centerline of the cross-section ($x=0$ or $y=0$ or neither) but not both. 
(v) Tumbling motion (purple) with trajectories that stay outside the central region of the cross-section as defined for central swinging, and wander near the channel walls. 
(vi) Wandering motion (red) with trajectories not in classes (i)--(v) and which, therefore, 
visit both the central region defined for central swinging motion, as well as the outer region defined for tumbling motion. Numerically, the classification is implemented by only analyzing the latter half of the trajectory to remove any transient dynamical behaviors at short times.


\subsection{Comparison with cylindrical channel}

For a square cross-section ($AR=1$) we have the approximate dimensionless fluid velocity from Eq.~\eqref{eq: flow field dimless},
$${u}_{s}(x,y)={U}(1-x^2)(1-y^2)={U}\left(1-(x^2+y^2)+x^2y^2\right).$$
By introducing the following velocity field
$${u}_{cs}(x,y)={U}\left(1-(x^2+y^2)+\alpha x^2y^2\right),\ \alpha\in[0,1],$$
we can continuously transform from the Poiseuille flow of a circular cross-section as in Eq.~\eqref{eq: flow circular} $(\alpha=0)$ to the approximate Poiseuille flow of a square cross-section ${u}_{s}(x,y)$ $(\alpha=1)$. 
The corresponding nonlinear ODEs for an active particle within the flow field ${u}_{cs}(x,y)$ are:
\begin{subequations}\label{eq: microswimmer general circle square}
\begin{align}
    \dot{x}&=-\cos\theta \sin\phi\\ 
    \dot{y}&=\sin\theta\\ 
    \dot{\theta}&=-{U}y\cos\phi + \alpha {U}x^2 y\cos\phi \\ 
    \dot{\phi}&={U}\left(-y\tan\theta \sin\phi  + x\right) \\
    &\quad+ \alpha {U} xy \left(x\tan\theta \sin\phi  -y\right) .\nonumber
\end{align}
\end{subequations}
It can be seen from Eq.~\eqref{eq: microswimmer general circle square} that the equilibrium states and their linear stability, are the same for a square cross-section (refer Sec.~\ref{sec: eq lin stab} for $AR=1$) 
and a circular cross-section~\citep{Zottl2012}. Specifically, since the $\alpha$ terms present in the $\dot{\theta}$ and $\dot{\phi}$ components of \eqref{eq: microswimmer general circle square} contribute at an order $O(\epsilon^3)$ near the equilibrium points their effects are not felt at order $O(\epsilon)$.


For our square channel, we only have the constant of motion $H_g$ in Eq.~\eqref{eq: COM general spehrical} (i.e. having implicitly utilised Eq.~\eqref{eq: COM trivial}). For ease of comparison with the quantity used by \citet{Zottl2012} for a circular cross-section, $H_c$ of Eq.~\eqref{eq: COM circle 1}, 
we define $H_s=1+\frac{1}{2}U+H_g\ge 0$,
and express the constant of motion here as
$$H_s=\frac{1}{2}{U}(x^2+y^2-x^2y^2)+1-\cos\phi \cos\theta.$$
Furthermore, the time derivative of the second constant of motion $L_z$ for the cylindrical channel in Eq.~\eqref{eq: COM circle 2},  has the following form for a square cross-section ($\alpha=1$):
\begin{equation}\label{eq: dLdt}
\frac{d L_z}{dt} = {U} x y \cos\theta \cos\phi \left(x^2-y^2\right).    
\end{equation}

We see that this quantity will not vary significantly when either $x$ or $y$ (or both) are small, or when $y\approx\pm x$.
Thus, for motion in a square cross-section confined near the channel center, along $x$ or $y$ axis or along diagonals, we expect the dynamics of the system to be regular (periodic/quasi-periodic) since the system is close to the cylindrical channel system which is both Hamiltonian and integrable. The central, vertical and horizontal swinging motions shown in Fig.~\ref{Fig: types of traj} are examples of this type of motion. Conversely, for general motion that is not restricted to these above regions such that the particle explores regions away from the center of the channel cross-section, the system deviates from a Hamiltonian integrable system and chaotic motion may arise. Tumbling and wandering motions (see Fig.~\ref{Fig: types of traj}) are examples of this. A comparison of active particle dynamics between circular and square cross-sections for a typical tumbling motion with the same initial conditions $(x(0),y(0),\theta(0),\phi(0))=(0.55,0.58,\pi/24,\pi/24)$ is shown in supplemental video S7.

\begin{figure*}
\centering
\includegraphics[width=1.8\columnwidth]{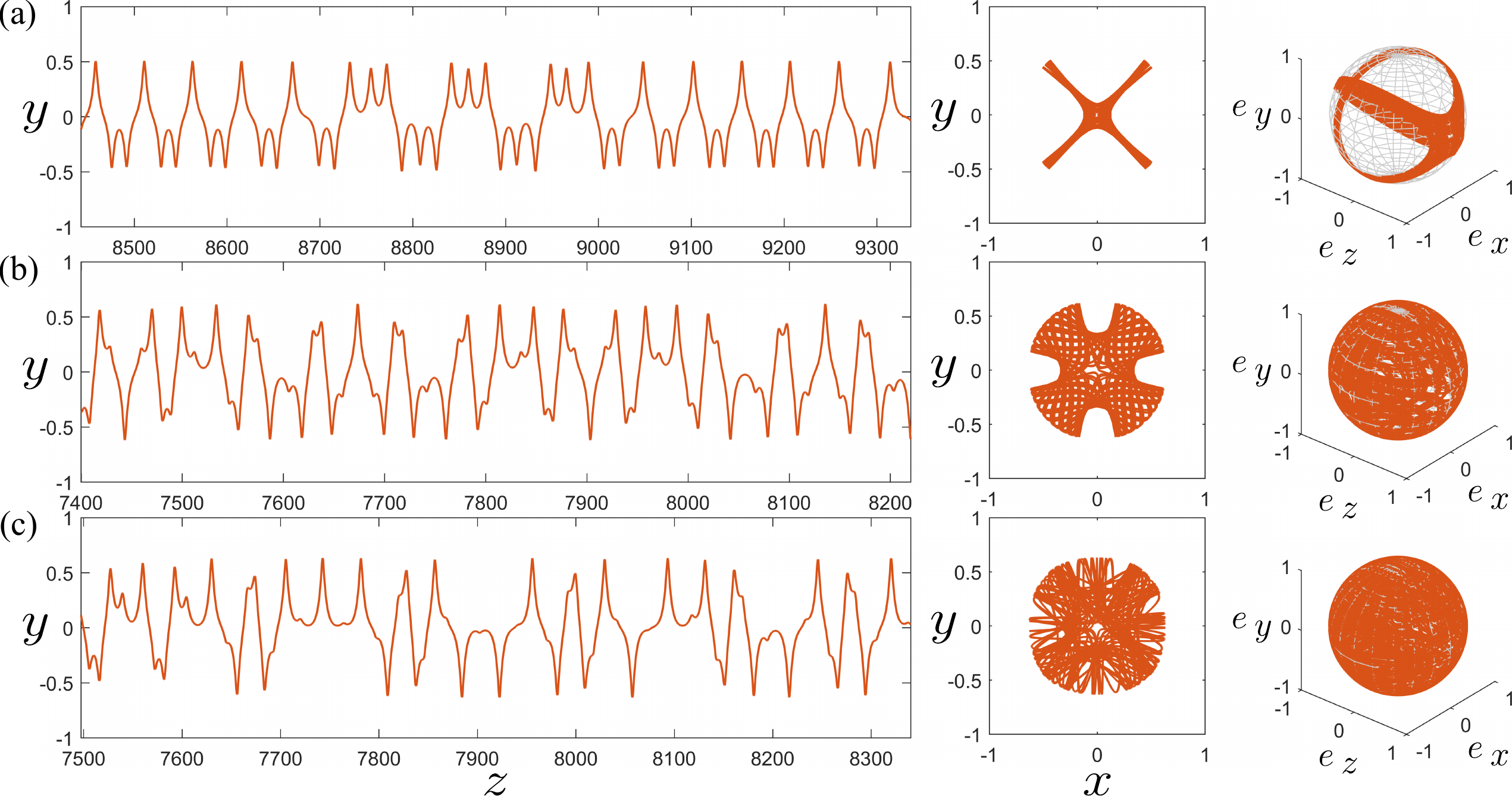}
\caption{Active particle motion for small perturbations near the unstable saddle equilibrium. Different trajectories are shown 
in the $(z,y)$ plane and $(x,y)$ plane for small perturbations (a) in the direction of the unstable manifold $(x(0),y(0),\theta(0),\phi(0))=(0,0,0,\pi) + (0.02,0.02,0.02\sqrt{{U}},0.02\sqrt{{U}})$, (b) in the direction of the stable manifold  $(x(0),y(0),\theta(0),\phi(0))=(0,0,0,\pi) +(0.02,0.02,-0.02\sqrt{{U}},-0.02\sqrt{{U}})$ and (c) a small general perturbation $(x(0),y(0),\theta(0),\phi(0))=(0,0,0,\pi) +(0.01,0.02,0.015,-0.005)$. The parameter ${U}=10$ was fixed and $z(0)=0$. See also supplemental videos S8-S10 for videos of panels (a)-(c), respectively.}
\label{Fig: traj near unstable equilibria}
\end{figure*}

\subsection{\label{sec: motion eq} Motion near equilibrium states}

Since the equilibrium states of the particle-fluid system are located at the center of the cross-section $(x=0,y=0)$ and the additional nonlinear $x^2 y^2$ term in the square channel flow field is small near these equilibrium points, we expect the motion near the equilibrium states to be similar to that of the circular cross-section~\citep{Zottl2012}. 

The eigenvectors corresponding to eigenvalues $\lambda_{1,2}=\pm i \sqrt{U}$ for the upstream-oriented particle equilibrium are
\begin{gather*}
\mathbf{v}_{1} = a_1 
\begin{bmatrix} 
 1 \\
 0 \\
 0 \\
 - i \sqrt{{U}} 
 \end{bmatrix}
+ b_1
\begin{bmatrix} 
 0 \\
 1 \\
 i \sqrt{{U}} \\
 0 
 \end{bmatrix}
\end{gather*}
and 
\begin{gather*}
\mathbf{v}_{2} = a_2 
\begin{bmatrix} 
 1 \\
 0 \\
 0 \\
 i \sqrt{{U}} 
 \end{bmatrix}
+ b_2
\begin{bmatrix} 
 0 \\
 1 \\
 -i \sqrt{{U}} \\
 0 
 \end{bmatrix},
\end{gather*}
respectively. Here, $a_1,b_1,a_2$ and $b_2$ are complex constants with $\bar{a}_1=a_2$ and $\bar{b}_1=b_2$, where the overbar denotes the complex conjugate. For small perturbations around this equilibrium point, we numerically observe periodic and quasiperiodic active particle motion confined near the center of the channel~(e.g. see Fig.~\ref{Fig: types of traj}(a) and supplemental video S1). We further find that the motion decouples in the $(x,\phi)$ and $(y,\theta)$ variables near this equilibrium point and we obtain the following system for the linearized equations of motion: 
\begin{align*}
    &\ddot{\theta}_1+{U}\theta_1=0,\\ 
    &\dot{y}_1= \theta_1,\\
    &\ddot{\phi}_1+{U}\phi_1=0,\\ 
    &\dot{x}_1 = -\phi_1.
\end{align*}
Thus, the evolution of active particle orientations $\theta$ and $\phi$ follow simple harmonic motion with oscillating frequency $\sqrt{{U}}$ and these oscillating orientations drive the translational motion of the active particle near this equilibrium point. Hence, the response to general small perturbations around this equilibrium point is a superposition of the above two decoupled oscillatory motions.

The eigenvectors corresponding to eigenvalues $\lambda_{3,4}=\pm \sqrt{{U}}$ for the downstream-oriented equilibrium point are 
\begin{gather*}
\mathbf{v}_{3} = a_3 
\begin{bmatrix} 
 1 \\
 0 \\
 0 \\
 \sqrt{{U}} 
 \end{bmatrix}
+ b_3
\begin{bmatrix} 
 0 \\
 1 \\
 \sqrt{{U}} \\
 0 
 \end{bmatrix}
\end{gather*}
and 
\begin{gather*}
\mathbf{v}_{4} = a_4 
\begin{bmatrix} 
 1 \\
 0 \\
 0 \\
 -\sqrt{{U}} 
 \end{bmatrix}
+ b_4
\begin{bmatrix} 
 0 \\
 1 \\
 -\sqrt{{U}} \\
 0 
 \end{bmatrix},
\end{gather*}
respectively, with real constants $a_3,b_3,a_4$ and $b_4$. This equilibrium is a saddle point with its unstable manifold tangent to the hyperplane spanned by the two basis vectors defining $\mathbf{v}_3$ while its stable manifold is tangent to the hyperplane spanned by the two basis vectors defining $\mathbf{v}_4$. To understand the nature of trajectories with small perturbations around this equilibrium point, we simulated motion with different initial perturbations. Some typical trajectories starting near this unstable equilibrium are shown in Fig.~\ref{Fig: traj near unstable equilibria} for perturbations in the directions of the stable/unstable manifolds as well as a general perturbation. We see that for a perturbation in the direction of the unstable manifold of the saddle, Fig.~\ref{Fig: traj near unstable equilibria}(a) and supplemental video S8, we obtain a cross-shaped trajectory in the channel cross-section that switches aperiodically between the four diagonals. The chaos here appears to be low-dimensional since the trajectory 
traces a well-defined path on the cross-shape with unpredictability only in the selection of the branches it traverses. For a perturbation in the direction of the stable manifold, Fig.~\ref{Fig: traj near unstable equilibria}(b) and supplemental video S9, we obtain a fan-shaped apparently quasiperiodic trajectory, while for a general perturbation as shown in Fig.~\ref{Fig: traj near unstable equilibria}(c) and supplemental video S10, we obtain a chaotic trajectory with no clear structure. In all cases, we find that a small perturbation from this unstable saddle equilibrium leads to wandering-type trajectories that move away from the equilibrium and explore both the inner and outer regions of the square cross-section. 


\subsection{\label{sec: PS} Exploration of the parameter space}

\begin{figure}
\centering
\includegraphics[width=\columnwidth]{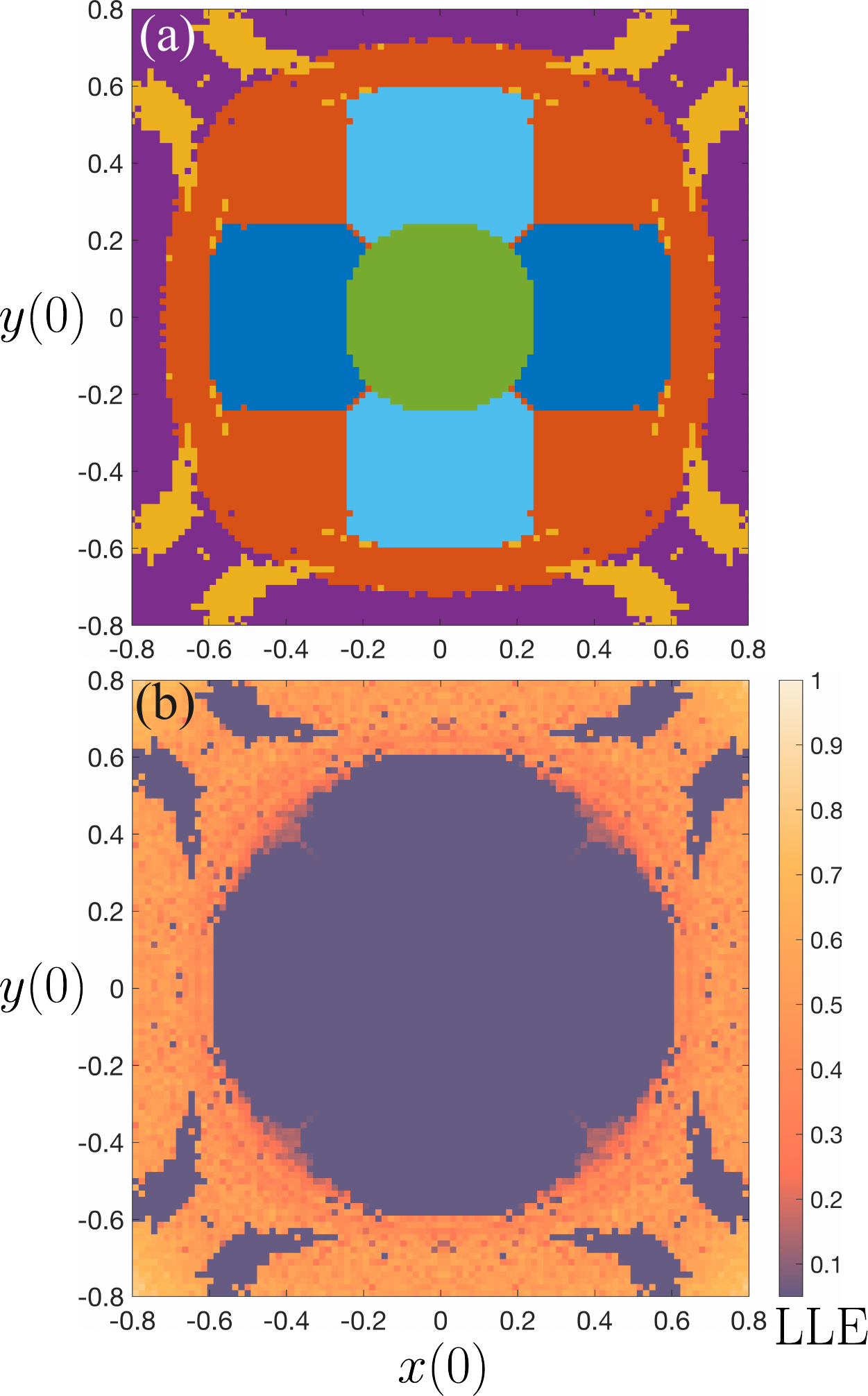}
\caption{(a) Classification of active particle trajectories and (b) largest Lyapunov exponent (LLE) in the initial position $(x(0),y(0))$ plane for the types of trajectories shown in Fig.~\ref{Fig: types of traj} with fixed ${U}=10$, $\theta(0)=0$ and $\phi(0)=0$. Green~(\protect\Mgreen) is central swinging motion, cyan~(\protect\Mcyan) is vertical swinging motion, blue~(\protect\Mblue) is horizontal swinging motion, yellow~(\protect\Myellow) is off-centered trapping motion, purple~(\protect\Mpurple) is tumbling motion and red~(\protect\Mred) is wandering motion.}
\label{Fig: Basin Lya square fixed theta phi}
\end{figure}

We have observed that both the initial conditions and the system parameters can greatly influence the type of active particle trajectory that is realized, and have classified these 
in Fig.~\ref{Fig: types of traj}. In this section, we explore the effects of the initial position and orientation, as well as the parameter ${U}$ on the active particle motion. 

The active particle dynamics are described by the $4$D nonlinear dynamical system in Eq.~\eqref{nonlinear eq AR} which requires four initial conditions: two position co-ordinates $x(0)$ and $y(0)$, and two orientation angles $\theta(0)$ and $\phi(0)$. To explore the solution space across the four initial conditions, we fix two of them and examine the types of active particle trajectories realized in the initial-condition plane formed by the remaining two. 

\subsubsection{\label{sec: initial position} Effect of initial position}

 We first fix the initial orientation of the active particle to point upstream, i.e. $\theta(0)=\phi(0)=0$, and explore the variation in active particle trajectories across different initial positions $(x(0),y(0))$. 
 We restrict the domain of initial positions to $x,y\in[-0.8,0.8]$. 
 This is done to exclude trajectories that get too close to the wall where interactions of the active particle with the wall may become important. 
 A plot depicting the classification of trajectories realized for different initial positions in the cross-section when ${U}=10$ is shown in Fig.~\ref{Fig: Basin Lya square fixed theta phi}(a). 
 We find that for an initial position near the center of the channel the motion remains confined near the center of the channel as indicated by the green region of Fig.~\ref{Fig: Basin Lya square fixed theta phi}(a), and that central swinging motion occurs~(see Fig.~\ref{Fig: types of traj}(a)). 
 An active particle starting out further away from the center of the channel but near to an axis remains confined near the same axis as indicated by the cyan and blue regions of of Fig.~\ref{Fig: Basin Lya square fixed theta phi}(a), which corresponds to vertical and horizontal swinging motions as in~Fig.~\ref{Fig: types of traj}(b,c). Along the diagonals and/or beyond the central region, Fig.~\ref{Fig: Basin Lya square fixed theta phi}(a) shows a red region corresponding to wandering trajectories~(see Fig.~\ref{Fig: types of traj}(f)). Near the edges of the square cross-section of Fig.~\ref{Fig: Basin Lya square fixed theta phi}(a), we have a purple region corresponding to tumbling trajectories~(as in Fig.~\ref{Fig: types of traj}(e)). Lastly, near the corners, Fig.~\ref{Fig: Basin Lya square fixed theta phi}(a) shows islands of yellow in the sea of purple corresponding to off-centered confined motion~(as in Fig.~\ref{Fig: types of traj}(d)). 

\begin{figure*}
\centering
\includegraphics[width=2\columnwidth]{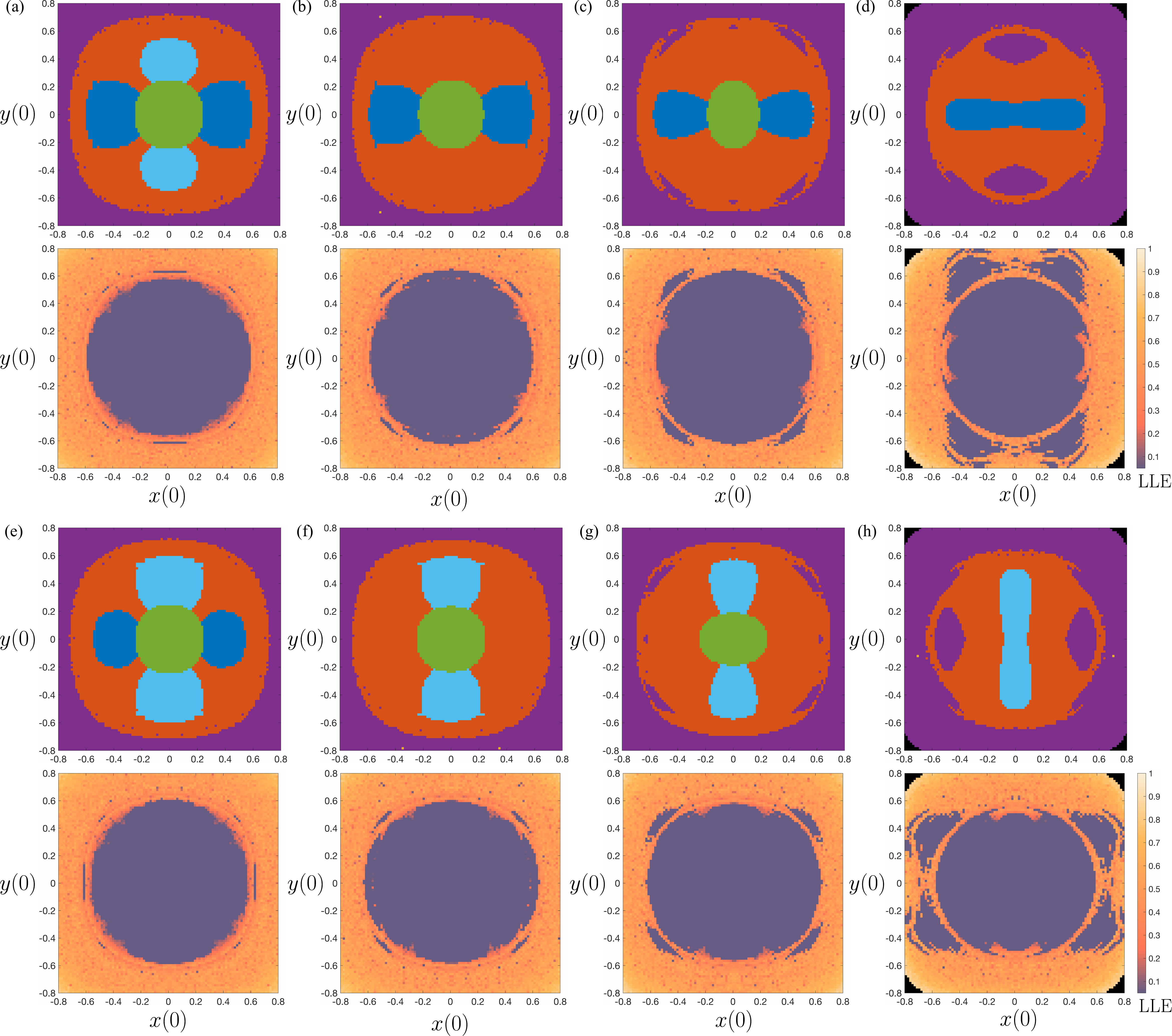}
\caption{Trajectory classification (top panels) and LLE (bottom panels) in the initial position $(x(0),y(0))$ plane for fixed ${U}=10$ and varying initial orientations $\theta(0)$ and $\phi(0)$. For fixed $\theta(0)=0$ and (a) $\phi(0)=\pi/24$, (b) $\phi(0)=\pi/12$, (c) $\phi(0)=\pi/6$ and (d) $\phi(0)=\pi/3$. For fixed $\phi(0)=0$ and (e) $\theta(0)=\pi/24$, (f) $\theta(0)=\pi/12$, (g) $\theta(0)=\pi/6$ and (h) $\theta(0)=\pi/3$. Green~(\protect\Mgreen) is central swinging motion, cyan~(\protect\Mcyan) is vertical swinging motion, blue~(\protect\Mblue) is horizontal swinging motion, purple~(\protect\Mpurple) is tumbling motion and red~(\protect\Mred) is wandering motion. Black regions in panels (d) and (h) correspond to trajectories that came too close to the channel walls i.e. trajectories that went outside a square of $[-0.95,0.95]\times[-0.95,0.95]$.}
\label{Fig: Basin square vary theta phi}
\end{figure*}

The above trajectory classification in the initial-condition space is based on the region occupied by the trajectory in the cross-section and does not necessarily capture information about the regular (periodic/quasiperiodic) or chaotic nature of trajectories. However, we typically find that trajectories which are confined near the center of the channel exhibit regular motion whereas active particles that travel near the edges of the cross-section show aperiodic motion and hints of chaos. Furthermore, the trajectories that travel near the edges can flip direction of motion around the origin between clockwise and counterclockwise~(e.g. see tumbling trajectory in supplemental videos S5 and S7). The presence of chaos in these trajectories can be quantified by calculating the largest Lyapunov exponent (LLE) of the underlying nonlinear dynamical system~\citep{strogatz}. If the LLE is zero, then the active particle motion is either periodic or quasiperiodic, whereas a positive LLE indicates chaos (with the degree of sensitivity to initial conditions given by the magnitude of the LLE). The magnitude of 
the LLE in the plane of initial conditions $(x(0),y(0))$ is shown in Fig~\ref{Fig: Basin Lya square fixed theta phi}(b) for the same domain of parameter values as Fig~\ref{Fig: Basin Lya square fixed theta phi}(a). We typically find that particles starting in the central region of the square cross-section have regular motion, whereas particles starting near the walls of the channel~(outer red and purple regions of Fig.~\ref{Fig: Basin Lya square fixed theta phi}(a)) are chaotic. However, we also find anomalous periodic regions near corners (within the chaotic sea); a particle starting in these small regions 
shows regular dynamics. These anomalous periodic regions correspond to the yellow region in Fig.~\ref{Fig: Basin Lya square fixed theta phi}(a) of off-centered trapping motion. 


\subsubsection{\label{sec: initial orientation} Effect of initial orientation}

We now explore how the classification 
of trajectories in the $(x(0),y(0))$ initial-position plane varies with small changes in the (fixed) initial orientation angles $\theta(0)$ and $\phi(0)$ from the upstream-orientation.

First, we explore the effect of variations in $\phi(0)$ for fixed $\theta(0)=0$. For a small positive value of $\phi(0)=\pi/24$, we find that the blue and cyan regions shrink marginally from those shown in Fig.~\ref{Fig: Basin Lya square fixed theta phi}(a) (where $\phi(0)=0$) to those of the 
top panel of Fig.~\ref{Fig: Basin square vary theta phi}(a). 
Further increasing to $\phi(0)=\pi/12$, we find that the cyan region corresponding to vertically swinging motion vanishes as shown in Fig.~\ref{Fig: Basin square vary theta phi}(b). 
This is because the value of $\phi(0)$ is large enough that, regardless of initial position, the active particle cannot remain confined to the vertical strip which classifies vertical swinging motion. 
Further increasing $\phi(0)$ to $\pi/6$ and $\pi/3$ leads to the shrinkage and disappearance, respectively, of the green region of central swinging motion~(see Fig.~\ref{Fig: Basin square vary theta phi}(c,d)). Again, with a large value of $\phi(0)$ the active particle is unable to remain confined near the channel center. 
Further, we see that parts of the red region of wandering motion are increasingly replaced by purple tumbling motion with trajectories largely confined near the channel walls. 
In terms of the chaotic nature of the trajectories, 
the bottom panels of Fig.~\ref{Fig: Basin square vary theta phi}(a)--(d) show a progressive vertical stretching of the regular region as $\phi(0)$ increases. 
However, structure indicating chaotic motion also persists within this vertical band of regular motion. The black regions near the corners of Fig.~\ref{Fig: Basin square vary theta phi}(d) correspond to trajectories that came too close to the channel walls, i.e. trajectories that went outside the square domain $(x,y)\in[-0.95,0.95]\times[-0.95,0.95]$. We note that although initial non-zero values of $\phi$ breaks symmetry in $x$ direction, we do largely see a persistent left-right symmetry in Fig.~\ref{Fig: Basin square vary theta phi}. This is probably due to fact that only long-time behavior is captured in the classification of trajectories.

We see a similar trend as 
$\theta(0)$ is increased for fixed $\phi(0)=0$, but with the blue regions vanishing instead of the cyan regions, see Fig.~\ref{Fig: Basin square vary theta phi}(e--h). 
The green region also  disappears for large $\theta(0)$. Even reasonably small but non-zero 
values of $\theta(0)$ cannot give rise to a horizontally swinging motion and large $\theta(0)$ also precludes trajectories confined to the central region. Again the presence of up-down symmetry for initial non-zero values of $\theta$ is probably due to fact that only long-time behavior is captured in the classification of trajectories

We note that the yellow periodic islands present near the corner of the square cross-section in Fig.~\ref{Fig: Basin Lya square fixed theta phi}(a) are not seen in~Fig.~\ref{Fig: Basin square vary theta phi} when one of the orientation angles is non-zero. 
Nevertheless, such islands do persist for sufficiently small non-zero values of both orientation angle.

\subsubsection{\label{sec: effects Ubar} Effect of ${U}$}

\begin{figure}
\centering
\includegraphics[width=0.9\columnwidth]{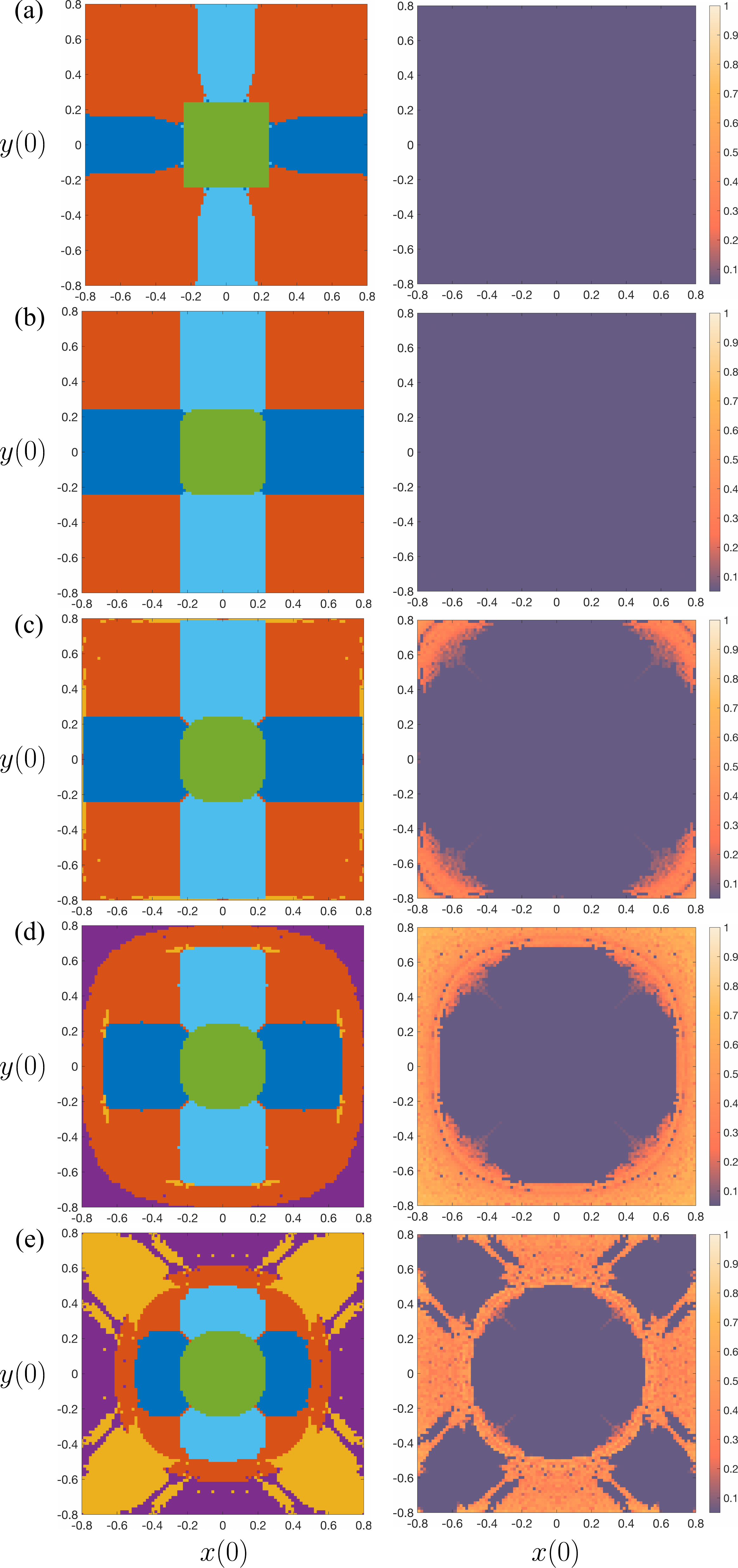}
\caption{Trajectory classification (left panels) and LLE (right panels) in the initial position $(x(0),y(0))$ plane for the types of trajectories shown in Fig.~\ref{Fig: types of traj} and different values of ${U}$: (a) ${U}=0.5$, (b) ${U}=2.5$, (c) ${U}=5$, (d) ${U}=7.5$ and (e) ${U}=15$. Initial orientation angles are fixed to $\theta(0)=0$ and $\phi(0)=0$. Green~(\protect\Mgreen) is central swinging motion, cyan~(\protect\Mcyan) is vertical swinging motion, blue~(\protect\Mblue) is horizontal swinging motion, yellow~(\protect\Myellow) is off-centered trapping motion, purple~(\protect\Mpurple) is tumbling motion and red~(\protect\Mred) is wandering motion.}
\label{Fig: Basin square vary U}
\end{figure}

\begin{figure*}
\centering
\includegraphics[width=1.75\columnwidth]{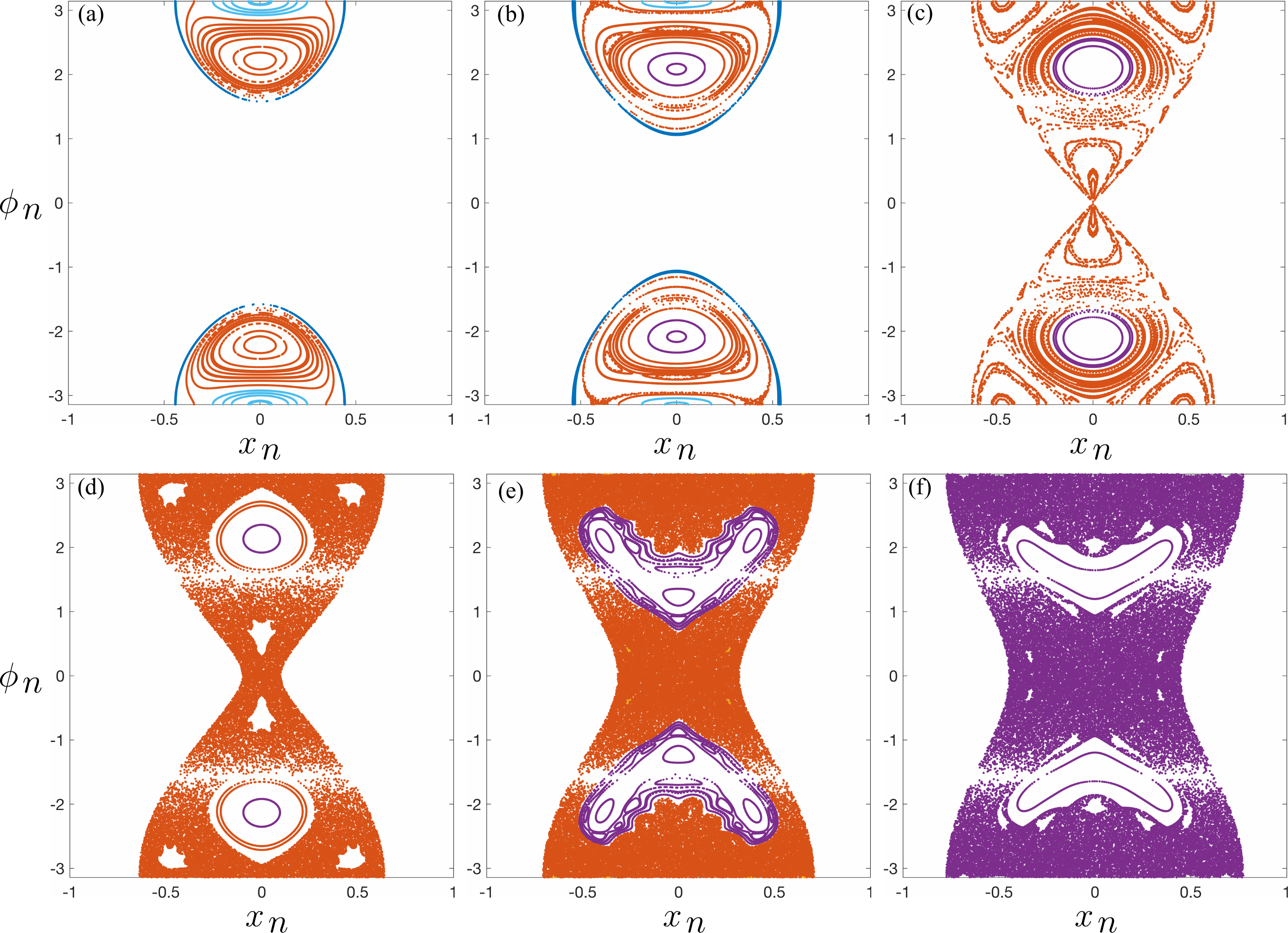}
\caption{Poincar\'e maps in the $(x_n,\phi_n)$ plane for $\theta(t_n)=0\:\:(\dot{\theta}(t_n)>0)$ and (a) $H_s=1$, (b) $H_s=1.5$, (c) $H_s=2$, (d) $H_s=2.05$, (e) $H_s=2.5$ and (f) $H_s=3$. The parameter ${U}=10$ was fixed and many different initial conditions were chosen based on Eq.~\eqref{eq: IC} while keeping $H_s$ fixed. Cyan~(\protect\Mcyan) is vertical swinging motion, blue~(\protect\Mblue) is horizontal swinging motion, purple~(\protect\Mpurple) is tumbling motion and red~(\protect\Mred) is wandering motion.}
\label{Fig: Poincare}
\end{figure*}

We now examine the variations in active particle dynamics with respect to the dimensionless parameter ${U}$, the ratio of the maximum flow speed to the intrinsic active particle speed. For ${U}\ll1$, the active particle intrinsic velocity dominates the flow field and hence, the particle will readily encounter the walls. Moreover, the active particle's self-generated flow fields would need to be captured to understand its dynamics. Since the simple model used in this work does not capture these two effects, we do not explore this regime of active particle motion in a quiescent fluid within a rectangular duct. 
For a recent detailed numerical exploration of this regime see \citet{activeparticlerect}. 

Figure~\ref{Fig: Basin square vary U} shows the different types of trajectories (left panels) and LLE (right panels) in the initial position space for various values of ${U}$ and fixed initial upstream orientation $\theta(0)=0$ and $\phi(0)=0$. For ${U}\lesssim1$, and as shown in
Fig.~\ref{Fig: Basin square vary U}(a) for $U=0.5$, we observe 
regular swinging motion of the active particle about the channel center. Moreover, at these low values of $U$ there is a net {\em upstream} migration of the particle, i.e. against the flow. 
For ${U}\gtrsim1$, and as shown in Fig.~\ref{Fig: Basin square vary U}(b) for $U=2.5$, we again observe regular swinging motion but now $U$ is sufficiently large that the net migration of the particle is downstream in the direction of the flow. Increasing to ${U}=5$, Fig.~\ref{Fig: Basin square vary U}(c), the classification of trajectories does not change qualitatively from that seen at lower values of $U$, 
but the plot of the LLE shows 
the emergence of chaotic motion for initial particle positions near the corners of the channel cross-section.
Further increasing ${U}$ to $7.5$, Fig.~\ref{Fig: Basin square vary U}(d), we see the emergence of purple regions of tumbling-motion trajectories near the walls of the channel 
where we also see an increase in the extent of chaotic regions. 
For a large value of ${U}=15$, Fig.~\ref{Fig: Basin square vary U}(e), we see the appearance of regular trajectories 
near the corners of the channel cross-section (yellow regions of the trajectory plot). 

To summarize, for the parameter values and the range of ${U}$ values shown in Fig.~\ref{Fig: Basin square vary U}, we find that as ${U}$ increases, the green region corresponding to central swinging motion does not change significantly while the cyan and blue regions of vertical and horizontal swinging shrink progressively. Further, with increasing ${U}$, the red region of wandering motion also shrinks with the appearance of tumbling and off-centered trapping motion near the edges and corners of the cross-section. In terms of the chaotic nature, we see a progressive increase in chaotic trajectories up to ${U}=7.5$, but further increase in ${U}$ leads to the appearance of regions of regular motion near the corners.

\begin{figure*}
\centering
\includegraphics[width=2\columnwidth]{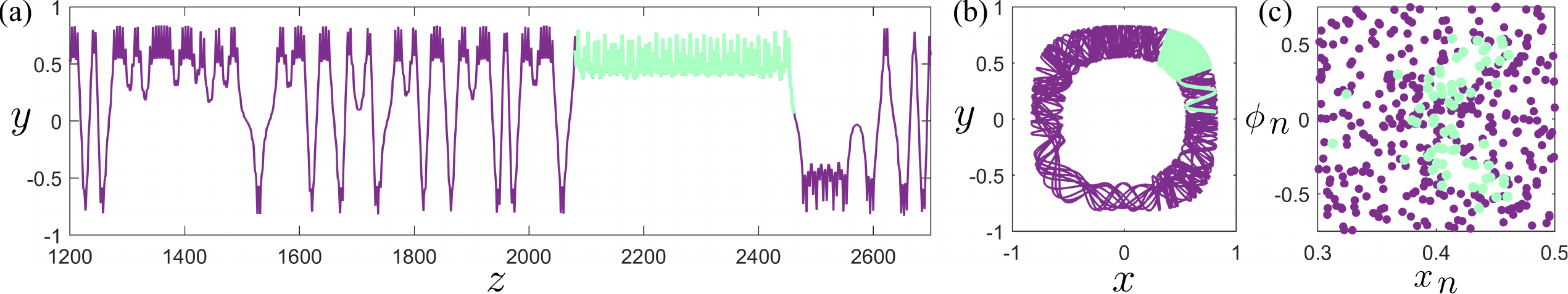}
\caption{Sticky trajectory for an active particle in a square cross-section. Particle trajectory in the (a) $(z,y)$ plane and (b) $(x,y)$ plane, as well as (c) Poincar\'e map in the $(x_n,\phi_n)$ plane for $\theta_n=0\,(\dot{\theta}_n>0)$. Purple shows the entire trajectory of the particle while light green shows a typical ``sticky" region for the trajectory. The parameter ${U}=10$ and initial conditions were chosen such that $H_s=3.5$ with $x(0)\approx0.6861$, $y(0)\approx0.6582$, $z(0)=0$, $\theta(0)=0$ and $\phi(0)=0$. See also supplemental video S11 for a video of the trajectory.}
\label{Fig: sticky trajetory}
\end{figure*}

\subsubsection{\label{sec: Poincare} Poincar\'e map and sticky trajectories}

Since one constant of motion $H_s$ remains for our $4$D dynamical system in Eq.~\eqref{nonlinear eq AR}, the effective dynamics of the system take place in $3$D. We can further explore the nature of the system dynamics and transition to chaos by using Poincar\'e sections to visualize regular and chaotic regions of the system on a $2$D plot. 
We construct a Poincar\'e map by sampling the active particle trajectory at times $t_n$ that correspond to a crossing of the phase-space trajectory with the $\theta=0$ (equivalently $e_y=0$) hyperplane in the positive direction, i.e. $\theta_n=\theta(t_n)=0$ with $\dot\theta_n=\dot{\theta}(t_n)>0$. 
At these times $t_n$, we store the values $\phi_n=\phi(t_n)$ and $x_n=x(t_n)$, and plot them against each other giving us a Poincar\'e map. We repeat this for many active particle trajectories having different initial conditions but a common fixed value of the constant of motion $H_s$, i.e. the initial conditions satisfy
\begin{align}\label{eq: IC}
H_s=&\frac{1}{2}{U}\left(x(0)^2+y(0)^2-x^2(0) y^2(0)\right)\\ \nonumber
&+1-\cos\left(\phi(0)\right)\, \cos\left(\theta(0)\right).
    \end{align}
Several such Poincar\'e maps are shown in Fig.~\ref{Fig: Poincare} for different values of the constant of motion $H_s$. We find that for small values of $H_s$, we observe regular behavior with ubiquitous quasiperiodic trajectories that correspond to closed curves on the Poincar\'e map~(see Fig.~\ref{Fig: Poincare}(a,b)). As $H_s$ is increased, we observe that, due to nonlinear resonances, some of these orbits break into a chain of smaller orbits~(see Fig.~\ref{Fig: Poincare}(c)). Further increase in $H_s$ gives rise to chaos as evident by the apparently random scatter of points in the Poincar\'e map~(see Fig.~\ref{Fig: Poincare}(d,e)). At these large values of $H_s$, we have a mixture of order and chaos where small islands of regular behaviors exist within the chaotic sea. 
Increase in the value of $H_s$ can also be interpreted as initial positions going away from the center of the channel. For example, with fixed initial upstream orientation $\theta(0)=\phi(0)=0$, Eq.~\eqref{eq: IC} describes a closed curve $$x(0)^2+y(0)^2-x^2(0)y^2(0)=2 H_s/{U}.$$
These closed curves are the same shape as the level sets of the flow field ${u}(x,y)$. Small values of $H_s$ correspond to circle-like closed curves near the center of the channel while increasing $H_s$ leads to square-like closed curves away from the center of the channel. Since near the center of the channel, the dynamics of the system are similar to a circular cross-section and hence integrable, this transition to chaos with increasing $H_s$ may be understood in terms of the theory of nearly integrable Hamiltonian systems and KAM theory~\citep{Zaslavsky2005-cl}.

When islands of regular behaviors exist within the chaotic sea of a Poincar\'e map, ``sticky" trajectories can arise where a long time is spend in the vicinity of these periodic islands~\citep{stickyorbits}. An example of such a sticky tumbling trajectory is shown in Fig.~\ref{Fig: sticky trajetory} and supplemental video S11. In this plot, panels (a) and (b) show the trajectory in different planes while panel (c) shows a Poincar\'e map with the light green colored part in all three panels representing the ``sticky" behavior. On the Poincar\'e map, such trajectories spend a very long time near the boundaries of the periodic islands compared to the time spent in a domain of the chaotic sea of the same phase-space volume.

\subsubsection{\label{sec: chaos} Dynamics in the large ${U}$ limit}

\begin{figure*}
\centering
\includegraphics[width=2\columnwidth]{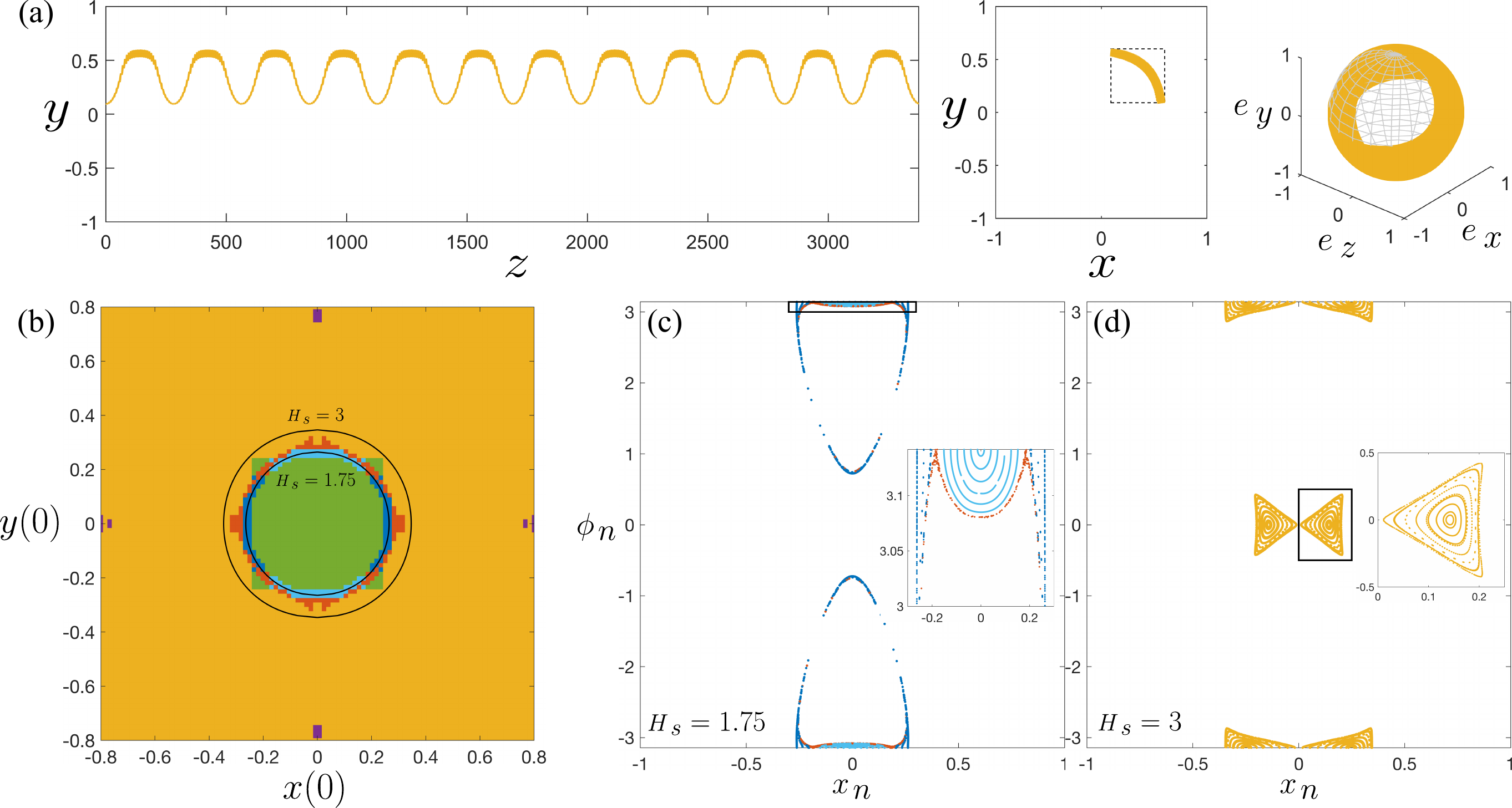}
\caption{Active particle dynamics for large ${U}=50$ starting at $z(0)=0$ with fixed initial upstream orientation angles $\theta(0)=0$ and $\phi(0)=0$. (a) A typical confined trajectory in the $(z,y)$ (left) and $(x,y)$ (middle) planes, as well as the orientation evolution (right) for initial position $x(0)=0.6,\ y(0)=0.1$. (b) Classification of behaviors in the $(x(0),y(0))$ initial-position plane using the color scheme of Fig.~\ref{Fig: types of traj}. (c,d) Poincar\'e maps in the $(x_n,\phi_n)$ plane for $\theta_n=0$ with $\dot{\theta}_n>0$ for the constants of motion (c) $H_s=1.75$ and (d) $H_s=3$. These values of the constant of motion are indicated by black curves in panel (b). The trajectory in panel (a) was simulated for $t=100$, using the time scaling of 
Eq.~\eqref{nonlinear eq AR}, which corresponds to $t=100\,{U}$ using the time scaling of 
Eq.~\eqref{nonlinear eq AR large U}. See also supplemental videos S12 for a video of the active particle trajectory in panel (a).}
\label{Fig: large U}
\end{figure*}

In the limit of large ${U}\gg 1$, Eq.~\eqref{nonlinear eq AR} approaches a singular limit for the evolution of $\theta$ and $\phi$. Hence, to understand this regime, we rescale the dimensionless time in Eq.~\eqref{nonlinear eq AR} (with $AR=1$) by ${U}$ to obtain the following system:
\begin{align}\label{nonlinear eq AR large U}
    \dot{x}&=-\frac{1}{{U}}\cos\theta \sin\phi\\ \nonumber
    \dot{y}&=\frac{1}{{U}}\sin\theta\\ \nonumber
    \dot{\theta}&=-y\cos\phi\left(1-x^2\right) \\ \nonumber
    \dot{\phi}&=-y\tan\theta \sin\phi \left(1-x^2\right) + x(1-y^2).
\end{align}
This form of the system removes the singular terms and allows efficient numerical solution. A typical classification of trajectories at a large value of ${U}=50$ along with an example trajectory and Poincar\'e maps are shown in Fig.~\ref{Fig: large U}. In this regime of large ${U}$ we find that chaos ceases and we have regular behaviors. A typical regular trajectory that confines itself to a quadrant of the channel is shown in Fig.~\ref{Fig: large U}(a). Simulating many different initial conditions with fixed initial upstream orientation $\theta(0)=\phi(0)=0$ gives us the classification in the initial position plane $(x(0),y(0))$ as shown in Fig.~\ref{Fig: large U}(b). Compared to Fig.~\ref{Fig: Basin square vary U}, we find that the central green region still exists while the cyan, blue and red regions have almost vanished. Outside the central region, the behavior is dominated by off-centered trapped trajectories (yellow region). Two different Poincar\'e maps at $H_s=1.75$ and $H_s=3$ are shown in Figs.~\ref{Fig: large U}(c) and (d) respectively. These typically show closed loops indicating quasiperiodic behavior of the system in this regime. With variation in the initial orientation angles $\theta(0)$ and $\phi(0)$ from the upstream equilibrium state, we find that in the initial-position plane, some yellow regions transition to purple regions corresponding to regular tumbling trajectories.

Hence, from Figs.~\ref{Fig: Basin square vary U} and \ref{Fig: large U} we see that the active particle motion is regular for small and large ${U}$, whereas chaos emerges for intermediate ${U}$.

\subsection{\label{sec: implications} Active particle transport along the channel}

\begin{figure*}
\centering
\includegraphics[width=2\columnwidth]{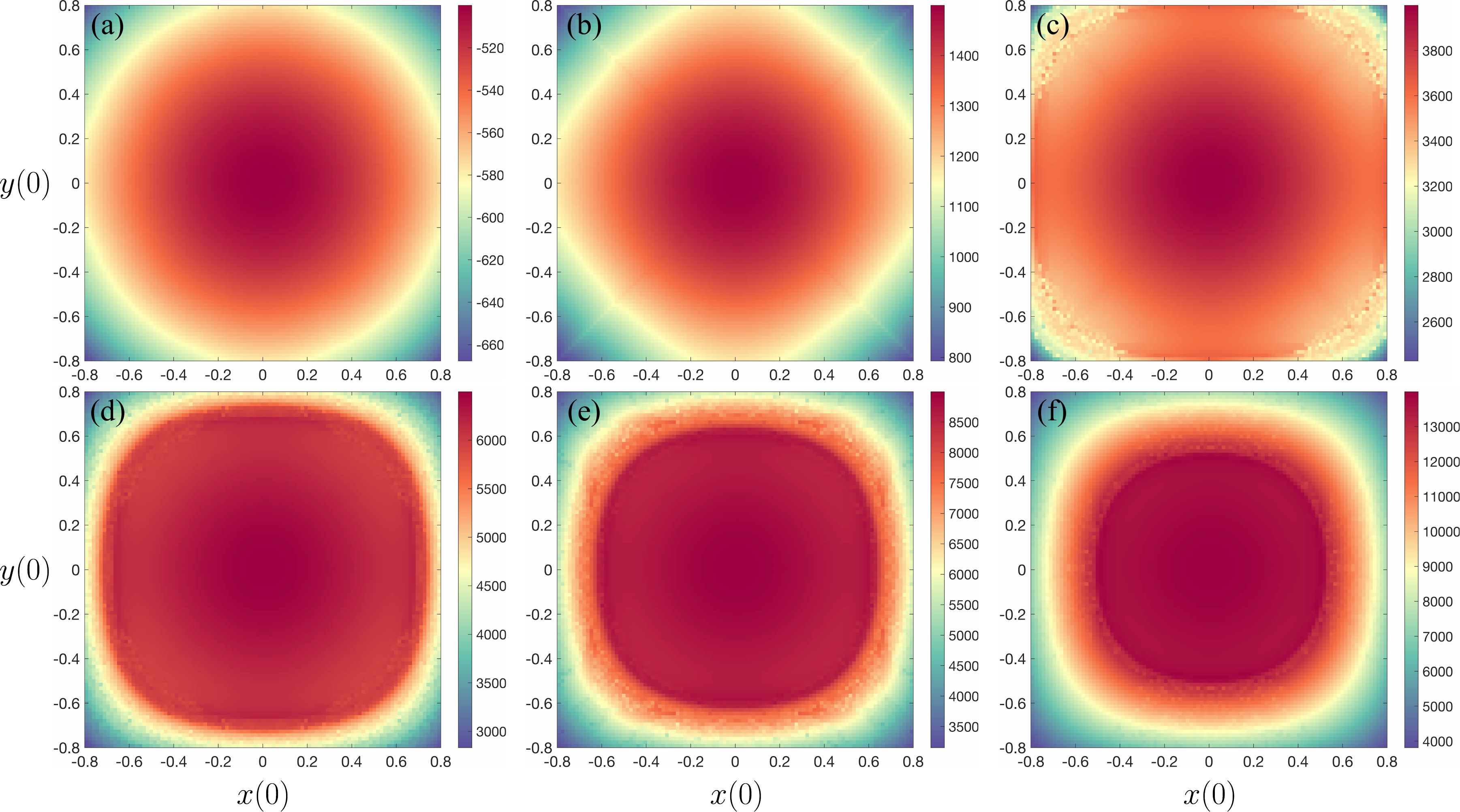}
\caption{Axial transport of the active particle along the straight channel based on the initial position in the $(x(0),y(0))$ plane for various ${U}$ and fixed $\theta(0)=0$ and $\phi(0)=0$. The colour bars indicate the final axial coordinate $z$ of the active particle from an initial axial position of $z(0)=0$ for values of (a) ${U}=0.5$, (b) ${U}=2.5$, (c) ${U}=5$, (d) ${U}=7.5$, (e) ${U}=10$ and (f) ${U}=15$.}
\label{Fig: z transport vary U}
\end{figure*}

The trajectories shown in Fig.~\ref{Fig: types of traj} started at $z=0$ and the time series is shown for $t=900$ to $t=1000$. We see that the transport of the active particle in the $z$ direction i.e. axially along with the flow, can vary significantly depending on the type of active particle motion realized in the channel cross-section. Active particles that perform swinging motion near the center of the channel~(e.g. green, blue, cyan and some red trajectories) will travel further along the channel compared to active particles whose motion is confined near the 
walls of the channel~(e.g. purple trajectories). Figure~\ref{Fig: z transport vary U} shows, for different ${U}$ values, a contour plot of the $z$ (axial) location at the end of the simulation ($t=1000$) in the $(x(0),y(0))$ plane for initially upstream-oriented active particles (starting from $z=0$). For a small value of ${U}=0.5$, we see that the final axial 
positions are negative indicating that the active particle's intrinsic speed dominates the fluid flow speed resulting in a net upstream migration of the active particle. For a larger value of ${U}=2.5$, the fluid speed dominates the particle speed and we obtain a net downstream migration. Further increase to ${U}=5$ does not qualitatively change the axial transport profile near the center of the cross-section, however, near the corners, we see fluctuations in this profile due to the appearance of chaotic wandering motion in this region~(see Fig.~\ref{Fig: Basin square vary U}(c)). For even larger values of ${U}$ the active particle axial transport is dominated by the background fluid flow profile. We find a central plug region corresponding to large axial transport of the active particle undergoing swinging motion near the center of the channel where flow speed is large, while near the walls and corners we observe small axial transport corresponding to off-centered trapping or tumbling trajectories that stay near the outer regions of the cross-section where the flow speed is small. 

\section{\label{sec: rect} Dynamics in wider rectangular channels}

\begin{figure*}
\centering
\includegraphics[width=2\columnwidth]{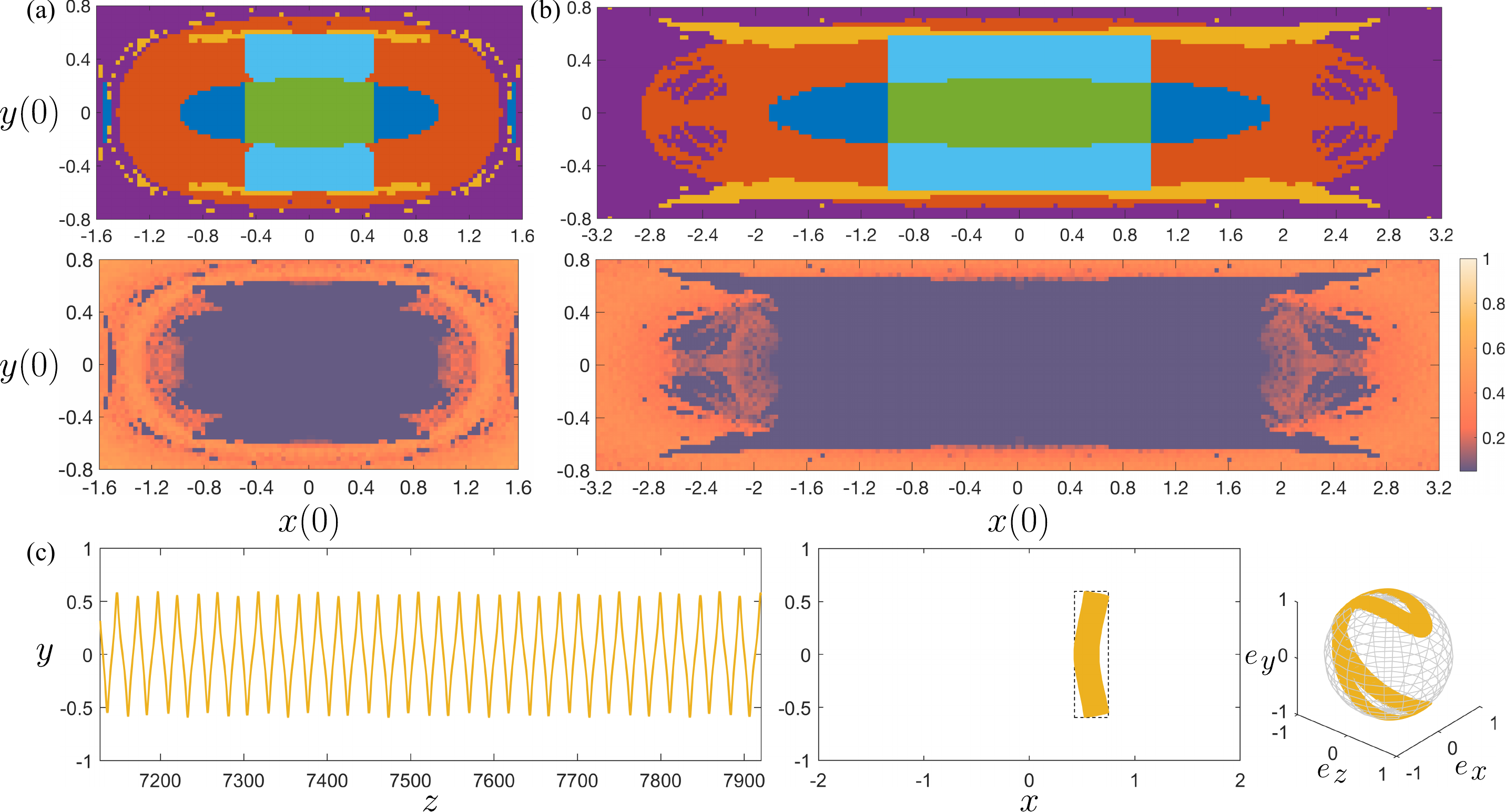}
\caption{(a,b) Active particle trajectory classification (top) and LLE (bottom) in the initial position plane $(x(0),y(0))$ for motion in rectangular channels with (a) $AR=2$ and (b) $AR=4$. (c) A typical trajectory of a particle starting in the yellow region with $x(0)=0.75$ and $y(0)=0.55$ for the case $AR=2$. Other parameters are fixed to ${U}=10$, $z(0)=0$, $\theta(0)=0$ and $\phi(0)=0$. See supplemental video S13 for a video of the active particle trajectory in panel (c).}
\label{Fig: rect channels}
\end{figure*}

In this section, we briefly explore the effects of the aspect ratio $AR$ of the rectangular cross-section on the active particle dynamics. In Fig.~\ref{Fig: rect channels}(a,b) we plot the trajectory classification and LLE in the initial-position plane $(x(0),y(0))$, while keeping $\theta(0)=0$, $\phi(0)=0$ and ${U}=10$ fixed, for two different rectangular cross-sections with $AR=2$ and $AR=4$, respectively. We note that the classification criteria presented for square channels in Sec.~\ref{sec: traj} has been scaled based on $AR$ in the $x$ direction for rectangular channels. So, for example, the classification of central swinging motion (green) has been modified to confinement in a rectangular box of domain $-0.25\,AR<x<0.25\,AR$ and $-0.25<y<0.25$. For the rectangular cross-section with $AR=2$, we find similar types of active particle trajectories at similar initial positions compared to the $AR=1$ square channel~(see Fig.~\ref{Fig: Basin Lya square fixed theta phi}) with some minor differences. Near the center of the channel, we obtain central swinging motion (green) as well as vertical (cyan) and horizontal (blue) swinging motion, but there is a horizontal stretching of these regions due to increasing the width of the channel. Moreover, the relative frequency of oscillations in the horizontal and vertical directions near the center of the channel will be scaled by $AR$ as per the eigenvalues in Sec.~\ref{sec: upstream eigenvalues}. Beyond the central region, we find wandering motion (red) and tumbling motion (purple) similar to the square channel. We find that the yellow islands, corresponding to confined trajectories away from the channel center, have diminished in size compared to the square cross-section, and these regions are more scattered. A typical trajectory in the yellow region is shown in Fig.~\ref{Fig: rect channels}(c) where the motion is trapped in an off-centered vertical band compared to the motion confined near the corner for a square cross-section~(see Fig.~\ref{Fig: types of traj}(d)). For the $AR=2$ channel, we also see the emergence of small horizontal swinging (blue) regions appearing near the left and right ends of the horizontal centerline that were not present for square channels. The LLE also shows similar structure with regular trajectories near the central region of the channel and the dominance of chaos near the channel walls. 
The $AR=4$ rectangular cross-section shows similar features to the $AR=2$ channel with a few noteworthy differences. The yellow regions for off-centered trapping motion increase in extent while the small regions for horizontal swinging present for $AR=2$ channel no longer exist. Moreover, the tumbling motion (purple) region penetrates the wandering motion (red) region near the left and right edges of the cross-section. We also note that our simple flow field approximation in Eq.~\eqref{eq: flow field dimless} will become poor near the left and right edges of the cross-section as $AR$ increases, and hence a more accurate flow field for Poiseuille flow in rectangular channels may be needed to accurately capture the active particle dynamics in these regions.

\section{\label{sec: concl} Conclusions}


We have studied in detail, theoretically and numerically, the motion of a point-like active particle in a steady unidirectional fluid flow, specifically through a straight channel with rectangular cross-section. We identified a general constant of motion that enabled the six equations of motion to be reduced to 
a $4$D nonlinear dynamical system with one constant of motion. We identified two equilibrium states for this particle-fluid system located at the center of the rectangular cross-section: (i) an upstream-oriented marginally stable equilibrium where small perturbations lead to oscillatory motion about this equilibrium point and (ii) a downstream-oriented unstable saddle equilibrium. By numerically solving the system, we observed a variety of active particle trajectories for different values of the maximum flow speed ${U}$ and the channel width/height aspect ratio $AR$, as well as different initial particle positions and orientations. The trajectories were classified based on the regions they occupy in the channel cross-section. Swinging trajectories, such as central swinging, vertical swinging and horizontal swinging, were typical 
quasiperiodic motions near the centerlines of the channel, whereas off-centered trapping motion 
was the typical form of confined quasiperiodic motion away from the channel centerlines. Tumbling trajectories stay near the walls of the channel 
while wandering trajectories visited both the central and the outer regions of the cross-section. By calculating the largest Lyapunov exponents, many of the tumbling and wandering trajectories were shown to be chaotic. Poincar\'e maps with increasing value of the constant of motion showed the transition to chaotic behavior and the persistence of small islands of regular behaviour in the chaotic sea. The latter resulted in ``sticky" chaotic tumbling trajectories due to the chaotic trajectory becoming trapped near a periodic state for a long time.

We have shown how the active particle motion varies with the system parameters and initial conditions.  We focused on a square channel cross-section ($AR=1$) and also showed that qualitatively similar particle trajectories were obtained in cross-sections with larger aspect ratio $AR$. 
Varying the maximum flow speed ${U}$, revealed rich dynamics with non-chaotic motion at very small and large ${U}$, and the emergence of chaos in an intermediate range of ${U}$. In this regime of ${U}$ where chaos arises, we found that the active particle trajectories are generally very sensitive to initial conditions with a couple of robust regimes. The active particle oriented upstream and starting near the channel center typically undergoes regular swinging motion that is robust to small variations in the initial position and orientation. Similarly, the active particle oriented upstream and starting near the walls of the channel typically undergoes chaotic tumbling motion which is again robust to small variations of the initial conditions. 

The model used in this paper is simple and can be extended in various ways to more accurately capture the motion of natural and artificial microswimmers in channel flows. The present model does not capture the interaction of the active particle with the channel walls and it would be useful to explore the effects of wall interactions to (i) understand motion of active particles in narrow channels and (ii) more accurately capture the active particle trajectories that get very close to the channel walls. The present work highlights the importance of initial conditions on active particle motion and hence it would be useful to perform careful microfluidic experiments to quantify the effects of initial conditions on active particle motion in channel flows. By considering wall interactions using squirmer models for the active particle in $2$D planar Poiseuille flow, \citet{Zottl2012} showed the emergence of dissipative dynamical features such as a stable point attractor and  a limit cycle attractor for the upstream-orientation swinging motion. \citet{Choudhary2022} explored the effects of adding fluid inertia for active particles in $2$D planar Poiseuille flow and reported similar dynamical features. Our previous work on the dynamics of passive spheres in $3$D channel flows with non-zero fluid inertia has revealed rich dynamical structure for inertial particle focusing behaviors~\citep{harding_stokes_bertozzi_2019,Valani2022SIADS,ValaniDSTA2021}. In future work, we aim to understand the effects of inertia on the dynamics and focusing of active particle in $3$D microfluidic channel flows.




\section*{Acknowledgments}
This research is supported under the Australian Research Council’s Discovery Projects funding scheme (project number DP200100834). Some of the results were computed using supercomputing resources provided by the Phoenix HPC service at the University of Adelaide.

\appendix

\begin{figure*}
\centering
\includegraphics[width=2\columnwidth]{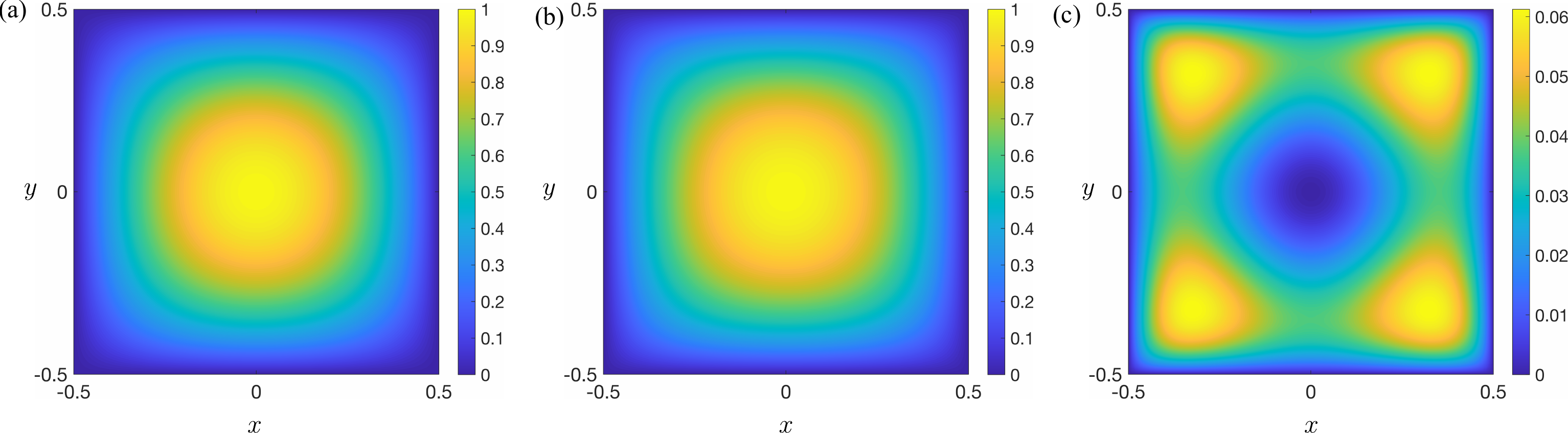}
\caption{Comparison of (a) the flow field ${u}(x,y)$ as in Eq.~\eqref{eq: flow field dimless} used in this paper to approximate Poiseuille flow in a $3$D rectangular channel (here setting $AR=1$ for a square channel) with (b) the more accurate representation of the flow field ${u}_{e}(x,y)$ as in Eq.~\eqref{eq: exact flow field} with $n=100$ terms.
The difference ${u}_{e}(x,y)-u(x,y)$ is shown in panel (c).}
\label{Fig: flow compare}
\end{figure*}

\begin{figure}
\centering
\includegraphics[width=0.75\columnwidth]{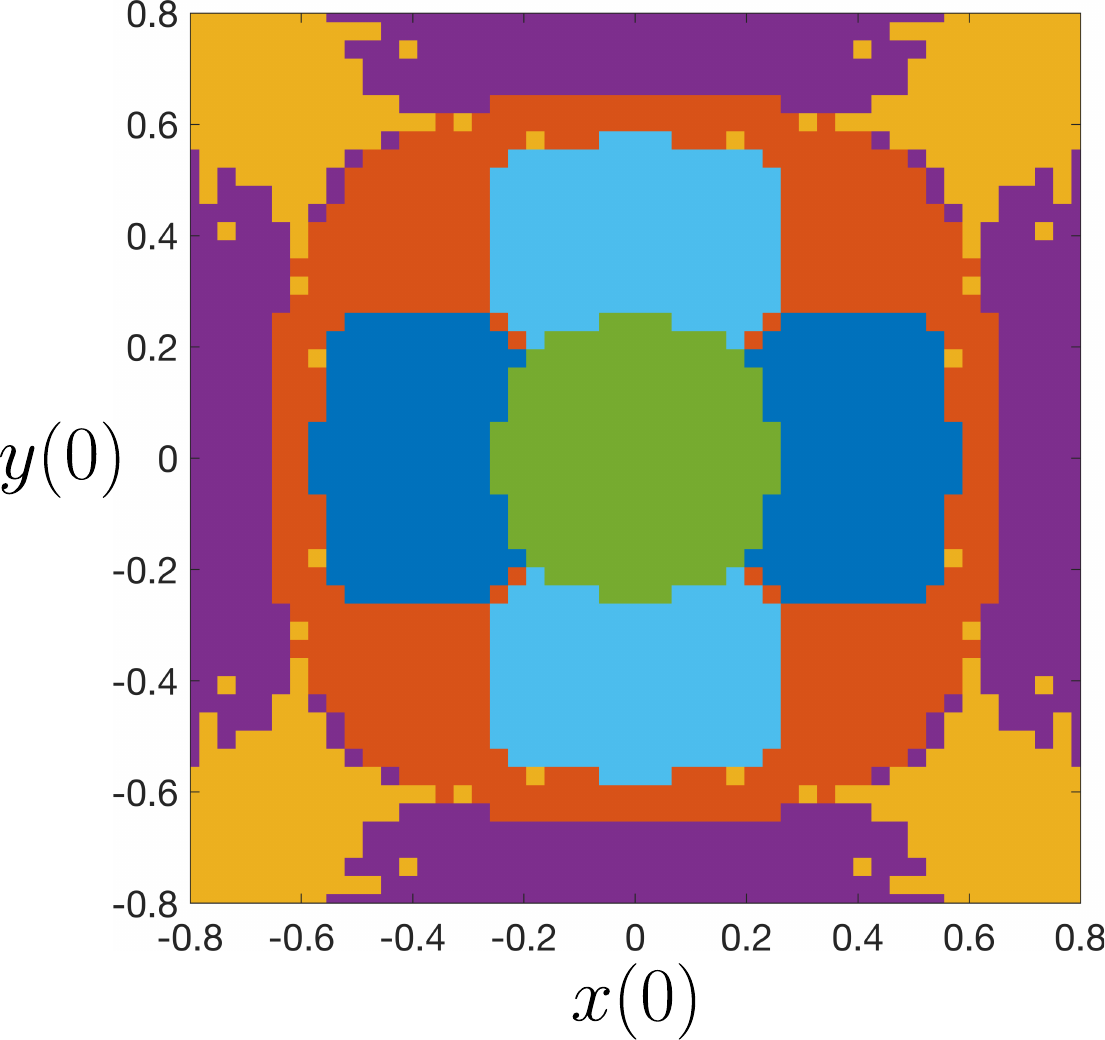}
\caption{Classification of active particle trajectories in the initial condition $(x(0),y(0))$ plane using the more accurate flow field in the square cross-section ${u}_{e}(x,y)$ (as per Eq.~\eqref{eq: exact flow field} with $n=100$ terms). This compares well with the classification shown in Fig.~\ref{Fig: Basin Lya square fixed theta phi}, obtained using the approximate flow field ${u}(x,y)$ with $AR=1$.
We note that the grid resolution here is $50\times50$ compared to the $100\times100$ resolution in Fig.~\ref{Fig: Basin Lya square fixed theta phi}. Other parameters were fixed to $\theta(0)=0$, $\phi(0)=0$ and $U=10$.}
\label{Fig: classificaiton exact flow}
\end{figure}

\section{Calculation for general constant of motion}\label{sec: general COM}

For dynamics of the active particle suspended in a unidirectional flow field ${u}(x,y)$ as given by Eqs.~\eqref{nonlinear eq AR cartesian full}, we show that following is a constant of motion:
$$H_g=-\frac{1}{2}{u}(x,y)+e_z.$$
Differentiating the above equation with respect to time we get
\begin{equation*}
 \frac{\text{d}H_g}{\text{d}t} = \frac{\partial H_g}{\partial x}\dot{x} + \frac{\partial H_g}{\partial y}\dot{y} + \frac{\partial H_g}{\partial z}\dot{z} + \frac{\partial H_g}{\partial e_x}\dot{e}_x + \frac{\partial H_g}{\partial e_y}\dot{e}_y + \frac{\partial H_g}{\partial e_z}\dot{e}_z.      
\end{equation*}
Calculating the derivatives and using Eq.~\eqref{nonlinear eq AR cartesian full} we get,
\begin{align*}
 \frac{\text{d}H_g}{\text{d}t} = -&\frac{1}{2}\frac{\partial {u}}{\partial x}e_x -\frac{1}{2}\frac{\partial {u}}{\partial y}e_y + 0 + 0 + 0\\ + &\left(\frac{1}{2}\frac{\partial {u}}{\partial x}e_x + \frac{1}{2}\frac{\partial {u}}{\partial y}e_y\right) = 0.
\end{align*}
Hence, $H_g$ is a constant of motion. If angular variables $\theta$ and $\phi$ are used for the particle orientation in place of $e_x, e_y$ and $e_z$, then the constant of motion transforms to
\begin{align*}
    H_g=-\frac{1}{2}{u}(x,y)-\cos\theta\cos\phi.
\end{align*}

\section{Comparison of exact versus approximate flow field for Poiseuille flow in a square cross-section}\label{sec: flow compare}

The dimensionless flow field ${u}(x,y)$ used in this paper is an approximation to the following exact flow field for Poiseuille flow in a straight duct with a square cross-section~\citep{Deville2022}:
\begin{align}\label{eq: exact flow field}
&{u}_e(x,y)={U}_e (1-y^2)\\ \nonumber
&+\frac{32{U}_e}{\pi^3}\sum_{n=0}^{\infty}\frac{(-1)^{n+1}\cosh\left({(2n+1)\frac{\pi x}{2}}\right)\,\cos\left({(2n+1)\frac{\pi y}{2}}\right)}{(2n+1)^3\,\cosh\left({(2n+1)\frac{\pi}{2}}\right)}.
\end{align}

Here ${U}_e={U}/\max\{{u}_e(x,y)\}={U}/{{u}_e(0,0)}$, to match the maximum flow speed of ${U}$ at the center of the channel. A comparison of the flow field ${u}$ and ${u}_e$ for ${U}=1$ is shown in Fig.~\ref{Fig: flow compare}. We see that the overall qualitative flow field is captured well by our approximate flow field ${u}$ and the difference between the two flow fields is small; at most $6\%$ when scaled by the maximum flow speed at the center. Further, we note that the regions in the cross-section where the most significant difference is observed in the flow field are near the corners of the square. For wider rectangular cross-sections, the approximation becomes poorer with increasing AR. For a $2\times1$ rectangular cross-section the maximum difference between the two velocity fields is around $15\%$ while for a $4\times1$ rectangular cross-section it is around $35\%$.

Figure~\ref{Fig: classificaiton exact flow} shows the different types of active particle trajectories realized for the more accurate flow field ${u}_e(x,y)$ for Poiseuille flow in a straight channel with square cross-section. Comparing with Fig.~\ref{Fig: Basin Lya square fixed theta phi}(a), which used the simpler approximation of the flow field ${u}(x,y)$, we find noticeable differences mainly in the yellow regions. This is to be expected since at these locations, the difference between the flow fields ${u}_e$ and ${u}$ is the largest~(see Fig.~\ref{Fig: flow compare}(c)).

\bibliography{apssamp}

\end{document}